\documentclass[a4paper,11pt]{article}
\pdfoutput=1
\usepackage{jheppub}

\usepackage{amssymb,amsmath}
\usepackage[normalem]{ulem}
\usepackage[utf8x]{inputenc}
\usepackage{slashed}
\usepackage{graphicx}
\usepackage{here}
\usepackage{color}
\usepackage{csquotes} 
\usepackage{comment}
\usepackage{mathrsfs}
\usepackage{float}
\usepackage{ascmac}
\usepackage{multirow}
\usepackage{longtable}

\usepackage[italicdiff]{physics}

\usepackage[hang,small,bf]{caption}
\usepackage[subrefformat=parens]{subcaption}
\usepackage{hyperref}

\usepackage{ulem}
\usepackage{xcolor}

\DeclareRobustCommand{\erase}{\bgroup\markoverwith{\textcolor{blue}{\rule[.5ex]{2pt}{0.4pt}}}\ULon}

\newcommand{\greekii}{I\hspace{-.1em}I}

\makeatletter
\newcommand*\rel@kern[1]{\kern#1\dimexpr\macc@kerna}
\newcommand*\widebar[1]{%
  \begingroup
  \def\mathaccent##1##2{%
    \rel@kern{0.8}%
    \overline{\rel@kern{-0.8}\macc@nucleus\rel@kern{0.2}}%
    \rel@kern{-0.2}%
  }%
  \macc@depth\@ne
  \let\math@bgroup\@empty \let\math@egroup\macc@set@skewchar
  \mathsurround\z@ \frozen@everymath{\mathgroup\macc@group\relax}%
  \macc@set@skewchar\relax
  \let\mathaccentV\macc@nested@a
  \macc@nested@a\relax111{#1}%
  \endgroup
}
\makeatother

\numberwithin{equation}{section}

\preprint{
\begin{minipage}{5cm}
\small
\flushright
EPHOU-25-002\\KEK-TH-2691\\KYUSHU-HET-311
\end{minipage}}

\title{Classification of Modular Symmetries in Type IIB Flux Landscape}

\author{Keiya Ishiguro$^{1}$,} 
\author{Takafumi Kai$^{2}$,}
\author{Tatsuo Kobayashi$^{3}$,} 
\author{Yuichi Koga$^{4}$, and}
\author{Hajime Otsuka$^{2}$} 
\affiliation{
$^1$KEK Theory Center, Institute of Particle and Nuclear Studies, KEK, 1-1 Oho, Tsukuba, Ibaraki 305-0801, Japan\\
$^2$Department of Physics, Kyushu University, 744 Motooka, Nishi-ku, Fukuoka 819-0395, Japan\\
$^3$Department of Physics, Hokkaido University, Sapporo 060-0810, Japan\\
$^4$Institute for Advanced Study, Kyushu University, 744 Motooka, Nishi-ku, Fukuoka 819-0395, Japan
}
\emailAdd{ishigu@post.kek.jp}
\emailAdd{kai.takafumi@phys.kyushu-u.ac.jp}
\emailAdd{kobayashi@particle.sci.hokudai.ac.jp}
\emailAdd{koga.yuichi.593@m.kyushu-u.ac.jp}
\emailAdd{otsuka.hajime@phys.kyushu-u.ac.jp}

\abstract{
In this work, we study modular symmetries in type IIB flux landscape by investigating symplectic basis transformations of period vectors on toroidal orbifolds.
To fix explicit cycles of a third-cohomology basis regarding the untwisted complex structure modulus, which is necessary to construct the period vectors, we find that the following two symmetries are required for the period vectors: (i) ``Scaling duality'' which is a generalized $S$-transformation of $PSL(2, \mathbb{Z})$ and (ii) the modular symmetries to be consistent with symmetries derived from mass spectra of the closed string in type IIB string theory.
Furthermore, by considering flux quanta on the cycles, we explore type IIB flux vacua on toroidal orientifolds and flux transformations under the modular symmetries of the period vectors.
}
\makeatletter
\gdef\@fpheader{}
\makeatother

\begin{document}

\maketitle

\section{Introduction}
\label{sec:intro}

String theory, which naturally describes gravity and quantum mechanics, is widely anticipated to provide the fundamental physical laws governing our Universe.
It was known that string theory admits enormous vacua in the context of flux compactifications.
Among those, the vacua arising from flux compactifications in type IIB string theory are referred to as type IIB flux vacua (for reviews, see \cite{Blumenhagen:2006ci, Grana:2005jc, McAllister:2023vgy}).
Since this landscape serves as one of the controllable low-energy effective theories derived from string theory, it has been extensively investigated from both theoretical and phenomenological perspectives.

\medskip

Modular symmetry in low-energy effective theories provides a powerful tool for clarifying the vacuum structure within the flux landscape.
In type IIB flux compactifications on $T^6/(\mathbb{Z}_2 \times \mathbb{Z}_2^\prime)$ orientifolds, the modular symmetry leads to a classification of Vacuum Expectation Values (VEVs) regarding complex structure moduli fields as physically independent vacua.
Then, the distribution of the VEVs was known to be clustered at fixed points of $PSL(2,\mathbb{Z})$ modular symmetry \cite{DeWolfe:2004ns,Ishiguro:2020tmo}.
In addition, although the modular symmetry appearing in the landscape concerning $T^6/\mathbb{Z}_{6-\rm \greekii}$ orientifold is not $PSL(2, \mathbb{Z})$ but $\bar{\Gamma}_0(3)$ which is the subgroup of $PSL(2, \mathbb{Z})$, it was found that the distribution of the VEVs is clustered at the fixed points of $\bar{\Gamma}_0(3)$ modular symmetry \cite{Ishiguro:2023flux}.
Thus, we can state that the flux landscape favors the fixed points of the modular symmetry.
From a phenomenological perspective, the fixed points of the modular symmetry are known to exhibit remarkably interesting features \cite{Ding_2019, Abbas_2021, Okada_2021, Feruglio_2021, Wang_2021, Ishiguro:2022pde, Kikuchi_2023, Knapp_P_rez_2023,Novichkov:2021evw,Petcov:2022fjf,Kikuchi:2023jap,deMedeirosVarzielas:2023crv}.
Therefore, it is very important to systematically classify the modular symmetries in low-energy effective theories to reveal the vacuum structure of the flux landscape in different toroidal orientifolds.

\medskip

Considering the systematic classification of the modular symmetries, it is useful to discuss the symmetries through the symplectic basis transformation of the period vector.
However, the methods mentioned in previous research \cite{Ishiguro:2023flux} are not enough to systematically classify the modular symmetries of the period vector in various toroidal orbifolds because we can not determine cycles of a third-cohomology basis on the toroidal orbifolds.
Since the linear combinations of these cycles lead to different cycles with different modular symmetries, we discuss the resolution of the indefiniteness of the cycles by choosing the period vector including the structure of ``Scaling duality'' which is a natural generalization of $S$-transformation of $PSL(2, \mathbb{Z})$.
In addition, we must consider the degree of freedom regarding overall factors of the cycles of the third-cohomology basis.
Since these factors affect the intersection numbers and the relative sizes between the components of the period vector, the indefiniteness of the overall factor causes the emergence of various modular symmetries.
Then, by assuming the consistency of duality symmetries for a complex-structure modulus on each toroidal orbifold in the context of low-energy effective action of string theory, we can fix the overall factors of cycles.
In heterotic string theory, the duality symmetries of the complex-structure modulus are derived from threshold corrections of moduli-dependent gauge coupling constants \cite{Mayr:1993mq, Bailin_1994}.
These previous researches have classified the duality symmetries by investigating the dependence of winding and momentum modes.
Therefore, since the structure of the winding and the momentum modes regarding the massive mode in the ten-dimensional bosonic string sector of the type IIB closed string is the same as the case of the heterotic string theory, we can classify the duality symmetries of the complex-structure modulus by applying the same technology to the one-loop partition function of the type IIB closed string.
Under this assumption, the systematic classification of the modular symmetries can be performed in the type IIB flux landscape of the various toroidal orientifolds. 

\medskip

This paper is organized as follows.
In section \ref{sec:period_geometry}, we briefly review the geometry of $T^6/\mathbb{Z}_4$ orbifold with $SU(2) \times SU(4) \times SO(5)$ root lattice.
First, we construct the cycles of the third cohomology basis which is invariant under the Coxeter element $Q$, and investigate these intersection numbers.
Next, we calculate complex one-forms by taking account of eigenvectors concerning the Coxeter element to obtain a holomorphic three-form.
In section \ref{sec:period_modular}, the construction of the period vector is discussed to study the modular transformation of it.
To include the contribution of the intersection numbers which are not normalized to one, we revisit the definition of the coordinates and the derivatives of a prepotential.
Moreover, we show the ``Scaling duality'' appearing in the symplectic transformation group $Sp(4, \mathbb{Z})$.
In addition, we summarize the duality symmetries on the toroidal orbifolds and mention the conditions for the period vector, which are necessary to classify the modular symmetries systematically.
In section \ref{sec:landscape}, by taking account of effective potentials induced by three-form fluxes on the toroidal orientifolds, we obtain supersymmetric (SUSY) Minkowski solutions by solving F-term equations. 
Then, to check the invariance of the F-term equations and the scalar potential under the modular transformations, we discuss the relevant transformations of the three-form fluxes.
Considering these flux transformations, we can deal with the explicit modular symmetries in the low-energy effective theory on each toroidal orientifold.
Section \ref{sec:conc} is devoted to the conclusion.
In Appendix \ref{sec:duality}, we discuss the details of the duality symmetries derived from the structure of the winding and the momentum numbers regarding the mass spectrum of the type IIB closed string.
In Appendix \ref{sec:orbifolds}, we summarize the explicit examples regarding the period vectors and the three-form fluxes on the various orientifolds.

\section{The geometry of orbifold}
\label{sec:period_geometry}

In the following, taking an explicit orbifold into account, we discuss the breakdown of the $SL(2, \mathbb{Z})$ symmetry for complex-structure moduli derived from the geometry of orbifolds with $h^{2,1}_{\rm untw.}=1$.
First, we perform $SL(2, \mathbb{Z})$ modular transformation of a period vector that appears in the effective action of type IIB string theory.
In particular, in this section, we focus on the $T^6/\mathbb{Z}_4$ orbifold with the $SU(2) \times SU(4) \times SO(5)$ root lattice.

\subsection{Geometry}
\label{sec:geometry}

The torus lattices for orbifold constructions and orbifold geometrical properties have been studied in \cite{Markushevich:1986za, Ibanez:1987pj, Katsuki:1989bf, Kobayashi:1991rp}.
In particular, the geometry of toroidal orbifolds utilized in flux compactifications of type IIB theory had been studied in \cite{Lust:2005dy, Lust:2006zg, reffert:2006to}, 
we briefly review them to facilitate later discussion.

First, we discuss the geometry of the $T^6/\mathbb{Z}_4$ orientifold with the $SU(2) \times SU(4) \times SO(5)$ root lattice.
The Hodge numbers for the untwisted sector are $(h^{1,1}_{\text{untw.}}, h^{2,1}_{\text{untw.}}) = (5, 1)$ and the Hodge numbers for the twisted sector are $(h^{1,1}_{\text{twist.}}, h^{2,1}_{\text{twist.}}) = (22, 2)$.
In this study, we focus exclusively on the untwisted sector of the complex-structure modulus and analyze the modular transformations that act on it.

To construct some orbifolds, we impose the boundary conditions on a basis of $T^6$.
When considering the $T^6/\mathbb{Z}_4$ orbifold, it is required that the six-dimensional metric is invariant under the action of $\mathbb{Z}_4$.
By using Coxeter element $Q$ corresponding to $SU(2) \times SU(4) \times SO(5)$ root lattice, we can define $\mathbb{Z}_4$ basis transformation for the metric described by $g_{ij} = \langle e_i, e_j \rangle$.
The action of $Q$ on $SU(2) \times SU(4) \times SO(5)$ root lattice is defined as
\begin{equation}
    \begin{aligned}
     Q(e_1) &= e_2, & Q(e_2) &= e_1 + e_2 + e_3 + e_4, \\
     Q(e_3) &= - e_1 - e_2 - e_3, & Q(e_4) &= - e_1 - e_2 - e_4, \\
     Q(e_5) &= e_6, & Q(e_6) &= - e_5 - e_6.
    \end{aligned}
\label{eq:twistactiononlattice}
\end{equation}
The matrix representation of $Q$ is defined by $Q(e_i) = e_j Q_{ji}$, and the explicit matrix is shown below:
\begin{align}
    Q = \begin{pmatrix}
        0 & 0 & -1 & 0 & 0 & 0 \\
        1 & 0 & -1 & 0 & 0 & 0 \\
        0 & 1 & -1 & 0 & 0 & 0 \\
        0 & 0 & 0 & 1 & -1 & 0 \\
        0 & 0 & 0 & 2 & -1 & 0 \\
        0 & 0 & 0 & 0 & 0 & -1 \\
    \end{pmatrix},
    \label{eq:matrixQ}
\end{align}
where $Q^4 = 1$.
From the invariance of the metric, i.e., $Q^t g Q = g$, and the relation between the metric and the 1-forms $dx^i$, i.e., $ds^2 = g_{ij} dx^i \otimes dx^j$, the 1-forms transform into $d\mathbf{x}' = Q d\mathbf{x}$.

In the case of $T^6$, there are 20 real three-forms defined by a third-cohomology basis as follows:
\begin{equation}
    \begin{alignedat}{5}
        \alpha_0 &= dx^1 \wedge dx^3 \wedge dx^5, \quad& \beta^0 &= dx^2 \wedge dx^4 \wedge dx^6, \\
        \alpha_1 &= dx^2 \wedge dx^3 \wedge dx^5, \quad& \beta^1 &= -dx^1 \wedge dx^4 \wedge dx^6, \\
        \alpha_2 &= dx^1 \wedge dx^4 \wedge dx^5, \quad& \beta^2 &= -dx^2 \wedge dx^3 \wedge dx^6, \\
        \alpha_3 &= dx^1 \wedge dx^3 \wedge dx^6, \quad& \beta^3 &= -dx^2 \wedge dx^4 \wedge dx^5, \\
        \\
        \gamma_1 &= dx^1 \wedge dx^2 \wedge dx^3, \quad& \delta^1 &= - dx^4 \wedge dx^5 \wedge dx^6, \\
        \gamma_2 &= dx^1 \wedge dx^2 \wedge dx^5, \quad& \delta^2 &= - dx^3 \wedge dx^4 \wedge dx^6, \\
        \gamma_3 &= dx^1 \wedge dx^3 \wedge dx^4, \quad& \delta^3 &= - dx^2 \wedge dx^5 \wedge dx^6, \\
        \gamma_4 &= dx^3 \wedge dx^4 \wedge dx^5, \quad& \delta^4 &= - dx^1 \wedge dx^2 \wedge dx^6, \\
        \gamma_5 &= dx^1 \wedge dx^5 \wedge dx^6, \quad& \delta^5 &= - dx^2 \wedge dx^3 \wedge dx^4, \\
        \gamma_6 &= dx^3 \wedge dx^5 \wedge dx^6, \quad& \delta^6 &= - dx^1 \wedge dx^2 \wedge dx^4,
    \end{alignedat}
    \label{eq:realbasis}
\end{equation}
where the six real coordinates $x^i$ on the torus $T^6$ have the periodic boundary conditions \textit{i.e.} $x^i \cong x^i + 1$ and the orientation is $\int_{T^6/\mathbb{Z}_4} dx^1 \wedge dx^2 \wedge dx^3 \wedge dx^4 \wedge dx^5 \wedge dx^6 = -1$.
Then, the above cycles satisfy
\begin{align}
    \int_{T^6/\mathbb{Z}_4} {\alpha_i \wedge \beta^j} = \delta^i_j, \quad \int_{T^6/\mathbb{Z}_4} {\gamma_i \wedge \delta^j} = \delta^i_j.
\end{align}
Note that the basis composed by these cycles is not the basis on $T^6/{\mathbb{Z}_4}$ but the basis on $T^6$ owe to lack of invariance for the Coxeter element and we just choose the normalization of intersection number regarding these cycles on $T^6/{\mathbb{Z}_4}$.

When discussing the geometry of orbifolds, it is necessary to prepare the real three-forms that are invariant under the Coxeter element $Q$ \cite{Ishiguro:2023flux}.
For example, by using a real three-form $\alpha_0$ and the definition of $Q^4 = 1$, we can obtain a new basis which is invariant under $Q$ as follows:
\begin{align}
   \begin{aligned}
        \sum \Gamma (\alpha_0) &:= Q (\alpha_0) + Q^2 (\alpha_0) + Q^3 (\alpha_0) + Q^4 (\alpha_0)\\
        &= 2 \alpha_0 - 2 \alpha_1 - 2 \delta^5 + 2 \delta^6,
   \end{aligned}
\end{align}
where we denote an orbit $\Gamma$ of $\alpha_0$ under a transformation group $G$ by $G (\alpha_0) \equiv \{ g \alpha_0 ~ | ~ g \in G \}$ and we consider the summation of each orbit regarding $\alpha_0$.
$Q$ is one of the elements in $G$ and the length of orbits regarding $\alpha_0$ is 4 due to $Q^4 (\alpha_0) = \alpha_0$.
Then, following the transformation: $d\mathbf{x}' = Q d\mathbf{x}$, we arrive at
\begin{align}
    \begin{aligned}
        Q (\alpha_0) &= Q (dx_1 \wedge dx_3 \wedge dx_5) \\
        &= (- dx_3) \wedge (dx_2 - dx_3) \wedge (2~ dx_4 - dx_5) \\
        &= - \alpha_1 - 2 \delta^5.
    \end{aligned}
\end{align}
To construct the cycles on the orbifold, we apply this action $\sum \Gamma$ to 20 real three-forms.
By choosing only the linearly independent cycles from the obtained invariant cycles, the following four cycles are derived
\footnote{Here, we can not fix the explicit basis by using the Coxeter element and intersection numbers.
In a later discussion, we choose the explicit form by modular symmetries originating from string theory.};
\begin{align}
    \begin{aligned}
    \mathbf{1}_{A^0} &\equiv \frac{1}{2} \sum \Gamma (\alpha_1), \\
    \mathbf{1}_{A^1} &\equiv \sum \Gamma (\beta^1), \\
    \mathbf{1}_{B_0} &\equiv \sum \Gamma (\delta^2), \\
    \mathbf{1}_{B_1} &\equiv \sum \Gamma (\alpha_0),
    \end{aligned}
    \label{eq:singletsasorbits}
\end{align}
which is invariant third-cohomology basis $H^3(T^6/{\mathbb{Z}_4}, \mathbb{Z})$ under the Coxeter element $Q$.
Here, we also consider the sum of orbits associated with the Coxeter element and the lengths of four orbits are
\begin{align}
    |\Gamma(\alpha_0)| = 4, \quad |\Gamma(\alpha_1)| = 4, \quad |\Gamma(\beta^1)| = 4, \quad |\Gamma(\delta^2)| = 4.
    \label{eq:lengthoforbit}
\end{align}
In terms of 20 real three-forms, the basis can be expressed as follows:
\begin{align}
    \begin{aligned}
    \mathbf{1}_{A^0} &= \alpha_0 - \gamma_2 - 2 \gamma_3 - \delta^5 - \delta^6, \\
    \mathbf{1}_{A^1} &= -2 \beta^0 - \gamma_5 + \gamma_6 - \delta^3, \\
    \mathbf{1}_{B_0} &= 2 \beta^1 + \gamma_5 - \gamma_6 - 2 \delta^2 - \delta^3, \\
    \mathbf{1}_{B_1} &= 2 (\alpha_0 - \alpha_1 -\delta^5 + \delta^6).
    \end{aligned}
    \label{eq:realorbits}
\end{align}
On $T^6$, these intersections of the dual cycles in regard to the orbifold satisfy the following relations:
\begin{align}
    \int_{T^6/\mathbb{Z}_4} \mathbf{1}_{A^0} \wedge \mathbf{1}_{B_0} = 4,\quad
    \int_{T^6/\mathbb{Z}_4} \mathbf{1}_{A^1} \wedge \mathbf{1}_{B_1} = 8,
    \label{eq:integralsofsinglets}
\end{align}
where the other intersection numbers are 0.
In section \ref{sec:conditions}, the difference of these intersection numbers is an important factor to obtain the modular transformation of the period vector, which is consistent with duality symmetries.

\subsection{Holomorphic three-form}
\label{sec:holomorphic_three-form}

For the construction of period vector which has the dependence of a complex-structure modulus, it is necessary to get the explicit expression of the holomorphic three-form.
A holomorphic three-form is defined as the wedge product of complex one-form $\{dz^i\}_{i=1,2,3}$ where the Coxeter element acts diagonally on these one-forms;
\begin{align}
    Q : \quad dz^i \rightarrow e^{2\pi \xi_i} dz^i,
\end{align}
with the eigenvalues $2\pi \xi_i$.
Then, in order to find the complex one-form, we require the following ansatz:
\begin{align}
    \begin{pmatrix}
        dz^1 \\
        dz^2 \\
        dz^3
    \end{pmatrix}
    =
    \begin{pmatrix}
        (v^1)^t d\mathbf{x} \\
        (v^2)^t d\mathbf{x}\\
        (v^3)^t d\mathbf{x}
    \end{pmatrix},
    \label{eq:definition_of_complex_coordinates}
\end{align}
where $v^i$ are the eigenvectors of the Coxeter element $Q^t$ with the eigenvalues $(e^{2\pi \xi_1}, e^{2\pi \xi_2}, e^{2\pi \xi_3})$ and the requirement of $SU(3)$-holonomy leads to $\pm \xi_1 \pm \xi_2 \pm \xi_3 =0$.
Here, we choose the eigenvalues $(\xi_1, \xi_2, \xi_3) = \left(\frac{1}{4}, \frac{1}{4}, -\frac{2}{4}\right)$.
In Eq. \eqref{eq:definition_of_complex_coordinates}, the action of the Coxeter element can be represented as
\begin{align}
    Q (dz^i) = (v^i)^t d\mathbf{x}' = (Q^t v^i)^t d\mathbf{x} = e^{2\pi \xi_i} dz^i.
\end{align}
From the above discussion, we can construct the complex one-forms as follows:
\begin{align}
    \begin{alignedat}{2}
    dz^1 &= a \left( dx^1 + i ~ dx^2 - dx^3 \right), \\
    dz^2 &= b \left( dx^4 - \left( \frac{1}{2} - \frac{i}{2} \right) dx^5 \right), \\
    dz^3 &= c \left( dx^1 - dx^2 + dx^3 \right) + d~ dx^6,
    \end{alignedat}
    \label{eq:pre_complexcoordinates}
\end{align}
where $a, b, c$ and $d$ are constants.

In the case of $T^6/\mathbb{Z}_4$, the volume of the fixed sub-torus regarding higher twists is larger than one \cite{reffert:2006to}.
By requiring the normalized periodicity for the sub-tori, we choose the normalization of Eq. \eqref{eq:pre_complexcoordinates} such that $|\text{det} Y| = 2 \text{Im} U$, where $Y$ is the transformation from real to complex coordinates. 

Since the direction of the fixed torus is derived from the invariance of the self-dual lattice \cite{Erler_1993}, we can describe the fixed torus by the winding number $w$ and momenta number $p$ satisfying $Q^2 w = w$ and $((Q^t)^{-1})^2 p = p$ (see Appendix \ref{sec:duality});
\begin{align}
    w = \begin{pmatrix}
        n_1 \\
        0   \\
        n_1 \\
        0   \\
        0   \\
        n_2 \\
    \end{pmatrix}, 
    \qquad
    p = \begin{pmatrix}
        m_1  \\
        -m_1 \\
        m_1  \\
        0    \\
        0    \\
        m_2  \\
    \end{pmatrix},
    \label{w&p}
\end{align}
where $\{n_1, n_2, m_1, m_2\}$ are integers.
The inner product of winding and momentum numbers is 
\begin{align}
    p^t w = 2 n^1 m^1 + n^2 m^2.
    \label{eq:volume_factor}
\end{align}
Then, the coefficient of $n^1m^1$, which is not one, implies that the real coordinate has a larger periodicity than one, and we consider a normalization to obtain the desirable periodicity in the following. 
Here, we apply Eq. \eqref{eq:volume_factor} to the eigenvectors of Eq. \eqref{eq:pre_complexcoordinates};
\begin{align}
    \begin{aligned}
        (v^1)^t w &= 0, \\
        (v^2)^t w &= 0, \\
        (v^3)^t w &= 2 c~ n_1 + d~ n_2.
     \end{aligned}
\end{align}
From this observation, we find that the fixed torus lies along the direction of $dz^3$ and we choose the normalization as $c = \frac{1}{2}$.
Finally, by interpreting the remaining complex degrees of freedom as the complex-structure modulus and considering $|\text{det} Y| = 2 \text{Im} U$, we can fix the complex one-forms as follows:
\begin{align}
    \begin{alignedat}{2}
    dz^1 &= \frac{1}{\sqrt{2}} \left( dx^1 + i ~ dx^2 - dx^3 \right), \\
    dz^2 &= dx^4 - \left( \frac{1}{2} - \frac{i}{2} \right) dx^5, \\
    dz^3 &= \frac{1}{2} \left( dx^1 - dx^2 + dx^3 \right) + U dx^6.
    \end{alignedat}
    \label{eq:complexcoordinates}
\end{align}
Consequently, by using these complex one-forms, the holomorphic three-form is defined as $\Omega = dz^1 \wedge dz^2 \wedge dz^3$ where $\int_{T^6/\mathbb{Z}_4} \Omega \wedge \bar{\Omega} = 2 i \text{Im} U$.

In the following section, we discuss the period vector which can be constructed by using the technique in \cite{Ishiguro:2023flux} and the basis transformation of the period vector is known to be associated with the subgroup of $PSL(2, \mathbb{Z})$.
However, the explicit form of the period vector is not fixed uniquely due to the indefiniteness of the third-cohomology basis on each orbifold.
In addition, it is found that the subgroup with respect to the transformation of the period vector changes when each component of the period vector changes.
Note that the complex one-forms are uniquely defined by taking account of $|\text{det} Y| = 2 \text{Im} U$ and the structure of the winding and the momentum numbers derived from the mass spectrum of the type IIB closed string, as discussed before. Given this fact, the modular symmetry of the period vector should be fixed uniquely because the duality symmetry for the complex-structure modulus on each toroidal orbifold is fixed uniquely.
Thus, it is strange that the modular transformation of the complex-structure modulus, which is derived from the basis transformation of the period vector, changes depending on how the third-cohomology basis is chosen.
To resolve this problem, we examine the consistency between the modular symmetry of the period vector and the duality symmetries in what follows.

\section{The modular transformation of period vector}
\label{sec:period_modular}

In this section, we discuss the construction of the period vector and the modular transformation. 
As mentioned in section \ref{sec:geometry}, the explicit real basis, which is invariant under the Coxeter element $Q$, can not be fixed and this result leads varieties of the modular transformation group regarding the period vector.
Our research takes account of determining the real basis by choosing one which is consistent with duality symmetries.

\subsection{Construction of period vector}
\label{sec:const_period}

To analyze the modular transformation of the complex-structure modulus, we focus on the period vector on $T^6/\mathbb{Z}_4$ orbifold which is defined by the $SU(2) \times SU(4) \times SO(5)$ root lattice.

Firstly, we introduce a basis of three cycles $A^I, B_J$ with $I, J = 0,..., h^{2,1}_{\rm untw.}$ on $T^6/\mathbb{Z}_4$ orbifold on general grounds \footnote{Here, these cycles are invariant under the Coxeter element $Q$ and we discuss only the untwisted modulus.}.
Then their intersection numbers are chosen as follows:
\begin{align}
    A^I \cap B_J = c _I\delta^I_J, \quad B_J \cap A^I = - c_I \delta^I_J, \quad A^I \cap A^I = 0, \quad B_J \cap B_J = 0, \qquad (c_I \in \mathbb{N}).
\end{align}
Note that we consider that these intersection numbers are not one.
The dual cohomology basis is expressed by $(\mathbf{1}_{A^I}, \mathbf{1}_{B_J})$;
\begin{align}
    \int_{A^J} \mathbf{1}_{A^I} = \int_{T^6/\mathbb{Z}_4} \mathbf{1}_{A^I} \wedge \mathbf{1}_{B_J} = c_I \delta^I_J, \qquad \int_{B_J} \mathbf{1}_{B_I} = \int_{T^6/\mathbb{Z}_4} \mathbf{1}_{B_I} \wedge \mathbf{1}_{A^J} = - c_J \delta^J_I. 
    \label{eq:gneral_intersection}
\end{align}
When the basis transformation preserves these properties, the group of transformation is described by a symplectic modular group $Sp(2 h^{2,1}_{\rm untw.} + 2, \mathbb{Z})$~\cite{Strominger:1990pd,Candelas:1990pi}.
Taking account of the definition of coordinates $X^I$ on the moduli space using $\mathbf{1}_{A^I}$ periods of holomorphic three-form and functions $F_I$ using $\mathbf{1}_{B_I}$ periods of holomorphic three-form, it is necessary to satisfy $\int_{T^6/\mathbb{Z}_4} \Omega \wedge \bar{\Omega} = 2 i \text{Im} U$.
In other words, we must consider the normalization of the holomorphic three-form on each cycle.
Thus, the coordinates $X^I$ and the functions $F_I$ are defined as follows 
\footnote{Relaxing the normalization for the intersection number ensures that the subsequent discussion demonstrates consistency between the string theory's duality symmetries and the period vector's modular symmetry.}:
\begin{align}
    X^I = \frac{1}{\sqrt{c_I}} \int_{A^I} \Omega, \qquad F_I = \frac{1}{\sqrt{c_I}} \int_{B_I} \Omega, \qquad (I = 0,..., h^{2,1}_{\rm untw.}).
    \label{eq:generalizedcoordinates}
\end{align}

In the case of $T^6/\mathbb{Z}_4$ orbifold, by using this definition, the holomorphic three-form can be expanded as
\begin{align}
    \Omega = \frac{1}{2} X^0 \mathbf{1}_{A^0} + \frac{1}{2 \sqrt{2}} X^1 \mathbf{1}_{A^1} - \frac{1}{2} F_0 \mathbf{1}_{B_0} - \frac{1}{2 \sqrt{2}} F_1 \mathbf{1}_{B_1}.
    \label{eq:expansion}
\end{align}
Applying this expansion Eq. \eqref{eq:expansion} to $\int_{T^6/\mathbb{Z}_4} \Omega \wedge \bar{\Omega}$, we can obtain an explicit expression of the period vector $\Pi$:
\begin{align}
    \begin{aligned}
        \int_{T^6/\mathbb{Z}_4} \Omega \wedge \bar{\Omega} &= - X^0 \bar{F}_0 - X^1 \bar{F}_1 + \bar{X}^0 F_0 + \bar{X}^1 F_1 \\
        &= \Pi^{\dagger} \Sigma \Pi,
    \end{aligned}
\end{align}
where
\begin{align}
    \Sigma = \begin{pmatrix}
        0 & \mathbf{1}_{2} \\
        - \mathbf{1}_{2} & 0
    \end{pmatrix}.
\end{align}
Therefore, we arrive at the following expression of the period vector
\begin{align}
    \begin{aligned}
        \Pi &\equiv \begin{pmatrix}
               X^0 \\
               X^1 \\
               F_0 \\
               F_1
               \end{pmatrix}
            &= \begin{pmatrix}
               \frac{1}{2} \int_{T^6/\mathbb{Z}_4} \Omega \wedge \mathbf{1}_{B_0} \\
               \frac{1}{2 \sqrt{2}} \int_{T^6/\mathbb{Z}_4} \Omega \wedge \mathbf{1}_{B_1} \\
               \frac{1}{2} \int_{T^6/\mathbb{Z}_4} \Omega \wedge \mathbf{1}_{A^0} \\
               \frac{1}{2 \sqrt{2}} \int_{T^6/\mathbb{Z}_4} \Omega \wedge \mathbf{1}_{A^1}
               \end{pmatrix} 
            &= \begin{pmatrix}
                \frac{1}{\sqrt{2}} \\
                -i U \\
                \frac{U}{\sqrt{2}} \\
                \frac{i}{2}
               \end{pmatrix},
    \end{aligned}
    \label{eq:period_vector}
\end{align}
which satisfies $\Pi^{\dagger} \Sigma \Pi = 2 i \text{Im} U$.
Since the overall factor can be absorbed by K\"{a}ler transformation, we consider a modified period vector $\Pi' = \left(1, -i\sqrt{2} U, U, \frac{i}{\sqrt{2}} \right)^t$ in the discussion of modular transformation.

\subsection{Modular transformation of period vector}
\label{sec:trans_period}

From the properties of Eq. \eqref{eq:gneral_intersection}, we know that the basis transformation of the period vector is $Sp(4, \mathbb{Z})$.
However, in general, we do not know how the symplectic transformation can be interpreted as a modular transformation of the complex-structure modulus in the period vector.\footnote{The relation between the modular transformation and symplectic transformation of the complex structure or K\"ahler structure moduli in specific Calabi-Yau threefolds was discussed in \cite{Ishiguro:2021ccl}.} 
In the following discussion, we take account of the modular transformation group regarding the period vector by a linear fractional transformation as follows:
\begin{align}
    U \rightarrow \frac{a U + b}{c U + d},
    \label{eq:lineartra}
\end{align}
where $a, b, c$ and $d$ are complex coefficients.
From the properties of Eq. \eqref{eq:gneral_intersection}, we know that the basis transformation group of the period vector is $Sp(4, \mathbb{Z})$.
Therefore, it is necessary to find constraints of $a, b, c$ and $d$ by making an association between the linear fractional transformation and the symplectic modular transformation.
The following discussion is based on \cite{Ishiguro:2023flux}.

First, regarding the period vector $\Pi' = \left(1, -i\sqrt{2} U, U, \frac{i}{\sqrt{2}} \right)^t$, we identify a relation between the linear fractional transformation with K\"{a}hler transformation $(c U + d)$ and the symplectic modular transformation.
Considering the transformation \eqref{eq:lineartra} for $\Pi'$, we can obtain the following vector
\begin{align}
    \begin{pmatrix}
        c U + d \\
        - i \sqrt{2} (a U + b) \\
        a U + b \\
        \frac{i}{\sqrt{2}} (c U + d)
    \end{pmatrix},
    \label{eq:modular_tra}
\end{align}
where $(c U + d)$ is absorbed by K\"{a}hler transformation.
In addition, the symplectic basis transformation of $\Pi'$ is described as follows:
\begin{align}
    X \Pi'
    = \begin{pmatrix}
        x_{11} - i \sqrt{2} U x_{12} + U x_{13} + \frac{i}{\sqrt{2}} x_{14} \\
        x_{21} - i \sqrt{2} U x_{22} + U x_{23} + \frac{i}{\sqrt{2}} x_{24} \\
        x_{31} - i \sqrt{2} U x_{32} + U x_{33} + \frac{i}{\sqrt{2}} x_{34} \\
        x_{41} - i \sqrt{2} U x_{42} + U x_{43} + \frac{i}{\sqrt{2}} x_{44} 
    \end{pmatrix},
    \label{eq:linear_symplectic}
\end{align}
where $X \equiv x_{ij}$ is generally chosen as an element of $Sp(4, \mathbb{Z})$.
In the following discussion, we consider the identification of these two vectors \eqref{eq:modular_tra} and \eqref{eq:linear_symplectic}.
Then, on both vectors, we take a quotient between the first and fourth components of the period vector:
\begin{align}
    - i \sqrt{2} = \frac{x_{11} - i \sqrt{2} U x_{12} + U x_{13} + \frac{i}{\sqrt{2}} x_{14}}{x_{41} - i \sqrt{2} U x_{42} + U x_{43} + \frac{i}{\sqrt{2}} x_{44}}.
    \label{eq:identity14}
\end{align}
Solving the above equation Eq. (\ref{eq:identity14}) as an identity, noting that $x_{ij}$ is an integer, yields following relations;
\begin{align}
    x_{14} &= -2 x_{41}, & x_{12} &= x_{43}, & x_{13} &= -2 x_{42}, & x_{11} &= x_{44}.
    \label{eq:conditions14}
\end{align}
For the second and third components of the period vector, similar calculations are performed and we can obtain the following relations:
\begin{align}
    x_{23} &= -2 x_{32}, & x_{21} &= x_{34}, & x_{24} &= -2 x_{31}, & x_{22} &= x_{33}.
    \label{eq:conditions23}
\end{align}
To satisfy the identities including Eq. \eqref{eq:identity14}, which is independent of the complex-structure modulus $U$, the matrix of symplectic transformation $X$ takes the form as follows:
\begin{align}
    X = \begin{pmatrix}
        x_{44} & x_{43} & -2 x_{42} & -2 x_{41} \\
        x_{34} & x_{33} & -2 x_{32} & -2 x_{31} \\
        x_{31} & x_{32} & x_{33} & x_{34} \\
        x_{41} & x_{42} & x_{43} & x_{44}
    \end{pmatrix}.
\end{align}
As a similar discussion, we take a quotient between the third and fourth components of the period vector on both vectors:
\begin{align}
    - i \sqrt{2} \frac{a U + b}{c U + d} = \frac{x_{31} - i \sqrt{2} U x_{32} + U x_{33} + \frac{i}{\sqrt{2}} x_{34}}{x_{41} - i \sqrt{2} U x_{42} + U x_{43} + \frac{i}{\sqrt{2}} x_{44}}.
    \label{eq:identity34}
\end{align}
By solving the above equation \eqref{eq:identity34} as an identity, additional conditions can be obtained as
\begin{align}
    \begin{aligned}
        x_{34} &= - \frac{2b}{c} x_{43}, & x_{41} &= \frac{d}{c} x_{43}, & x_{32} &= \frac{a}{c} x_{43}, \\
        x_{33} &= \frac{a}{d} x_{44}, & x_{42} &= - \frac{c}{2d} x_{44}, & x_{31} &= \frac{b}{d} x_{44}.
        \label{eq:conditions34}
    \end{aligned}
\end{align}
Taking account of conditions Eqs. \eqref{eq:conditions14}, \eqref{eq:conditions23} and \eqref{eq:conditions34}, we can obtain the matrix $X$ which is constrained by the identification of Eq. \eqref{eq:linear_symplectic}
\footnote{
The three conditions Eq. \eqref{eq:conditions14}, Eq. \eqref{eq:conditions23} and Eq. \eqref{eq:conditions34} are sufficient to construct the symplectic basis transformation matrix $X$.
When taking quotients between the first and second components, between the first and third components, and between the second and fourth components of the period vector, these conditions cannot restrict the matrix $X$ additionally, and then the symplectic basis transformation corresponds to the linear fractional transformation identically.
}:
\begin{align}
    X = \begin{pmatrix}
        x_{44} & x_{43} & \frac{c}{d} x_{44} & - \frac{2d}{c} x_{43} \\
        - \frac{2b}{c} x_{43} & \frac{a}{d} x_{44} & - \frac{2a}{c} x_{43} & - \frac{2b}{d} x_{44} \\
        \frac{b}{d} x_{44} & \frac{a}{c} x_{43} & \frac{a}{d} x_{44} & - \frac{2b}{c} x_{43} \\
        \frac{d}{c} x_{43} & - \frac{c}{2d} x_{44} & x_{43} & x_{44} \\
    \end{pmatrix}.
\end{align}
Considering a condition $X^t \Sigma X = \Sigma$ regarding the properties of the symplectic transformation, we find the following constraint:
\begin{align}
        (a d - b c) \left( \frac{2 x_{43}^2}{c^2} + \frac{x_{44}^2}{d^2} \right) = 1, 
\end{align}
and this equation can be rewritten as follows:
\begin{align}
    \begin{aligned}
        \frac{a d - b c}{d^2} \left\{ 2 \left(\frac{d}{c} x_{43} \right)^2 + x_{44}^2 \right\} &= 1, \\
        \frac{a}{d} x_{44} \cdot x_{44} - \frac{b}{d} x_{44} \cdot \frac{c}{d} x_{44} &= \frac{x_{44}^2}{2 \left( \frac{d}{c} x_{43} \right)^2 + x_{44}^2}.
    \end{aligned}
    \label{eq:constraintformatrix1}
\end{align}
Note that $X$ is the matrix of the symplectic transformation $Sp(4, \mathbb{Z})$ and these elements are integers.
Then, $\frac{a}{d} x_{44}$, $x_{44}$, $\frac{b}{d} x_{44}$ and $\frac{c}{d} x_{44}$ are the elements of the $Sp(4, \mathbb{Z})$ basis transformation matrix and LHS of Eq. \eqref{eq:constraintformatrix1} is an integer.
Therefore, an inequality $2 \left( \frac{d}{c} x_{43} \right)^2 + x_{44}^2 \leq x_{44}^2$ can be obtained, and we cannot consider the case of $x_{43} \neq 0$ and $x_{44} \neq 0$ because $\frac{d}{c} x_{43}$ is also the element of $X$ and an integer.
Here we can obtain the conditions $x_{43} = 0, x_{44} \neq 0$, and likewise, by taking account of the following equation:
\begin{align}
    \begin{aligned}
         - \left( - \frac{2a}{c} x_{43} \right) \cdot \frac{d}{c} x_{43} + \left( - \frac{2b}{c} x_{43} \right) \cdot x_{43} = \frac{2 x_{43}^2}{2 x_{43}^2 + \left( \frac{c}{d} x_{44} \right)^2},
    \end{aligned}
    \label{eq:constraintformatrix2}
\end{align}
we can obtain the conditions $x_{43} \neq 0, x_{44} = 0$.
Hence, we discuss two cases: Case I. $x_{43} = 0, x_{44} \neq 0$ and Case II. $x_{43} \neq 0, x_{44} = 0$.

\paragraph{Case I. $x_{43} = 0, x_{44} \neq 0$} \mbox{}\\

In this case, by using the condition $X^t \Sigma X = \Sigma$, an explicit matrix of $X$ is
\begin{align}
    X_1 = \frac{1}{\sqrt{a d - b c}}
    \begin{pmatrix}
    d & 0 & c & 0 \\
    0 & a & 0 & - 2 b \\
    b & 0 & a & 0 \\
    0 & - \frac{c}{2} & 0 & d \\
    \end{pmatrix}.
\end{align}
Since $X$ is the matrix of the symplectic transformation $Sp(4, \mathbb{Z})$ and these elements are integers, we must consider the following conditions for $a, b, c$ and $d$:
\begin{align}
    \begin{aligned}
        a' &\equiv \frac{a}{\sqrt{a d - b c}}, & b' &\equiv \frac{b}{\sqrt{a d - b c}}, \\
        c' &\equiv \frac{1}{2} \frac{c}{\sqrt{a d - b c}}, & d' &\equiv \frac{d}{\sqrt{a d - b c}},
    \end{aligned}
    \qquad (a', b', c', d' \in \mathbb{Z})
\end{align}
where these definitions lead $a' d' - 2 b' c' = 1$.
By using $a', b', c'$ and $d'$, we can describe the symplectic modular transformation as the restricted linear fractional transformation:
\begin{align}
    X_1 = \begin{pmatrix}
    d' & 0 & 2 c' & 0 \\
    0 & a' & 0 & - 2 b' \\
    b' & 0 & a' & 0 \\
    0 & - c' & 0 & d' \\
    \end{pmatrix}.
\end{align}
Thus, we conclude that the linear fractional transformation in the $x_{43} = 0$ case is Hecke congruence subgroup of level 2 denoted by $\Gamma_0(2)$;
\begin{align}
    \Gamma_0(2) \equiv \left\{\begin{pmatrix}
        a & b \\
        c & d
    \end{pmatrix} \in SL(2, \mathbb{Z}) \middle| ~ c \equiv 0 \quad \text{mod}~2
    \right\}.
\end{align}
This subgroup can be generated by generators $T$ and $V_1$ \cite{MR1027834} as follows:
\begin{align}
    T = \begin{pmatrix}
        1 & 1 \\
        0 & 1 \\
    \end{pmatrix}, \qquad
    V_1 = \begin{pmatrix}
        1 & 1 \\
        -2 & -1 
    \end{pmatrix},
\end{align}
where the elements $S$ and $T$ of $SL(2, \mathbb{Z})$ can describe generator $V_1$ as $S T S T^{-1} S$.
Therefore, we can find that the symplectic basis transformation for the period vector is consistent with the duality symmetry of the complex-structure modulus, as shown in Table \ref{tab:dualitygroups}.
Note that the modular transformation is not $SL(2,\mathbb{Z})$ but $PSL(2,\mathbb{Z})$.
Thus it is necessary to consider the corresponding subgroup $\bar{\Gamma}_0 (2)$ of $PSL(2, \mathbb{Z})$.\footnote{These modular subgroups can also lead to interesting modular flavor symmetries \cite{Kobayashi:2024ysa}.}
In general, the subgroup $\bar{\Gamma}_0(n)$ is defined as follows:
\begin{align}
    \bar{\Gamma}_0(n) \equiv \left\{\begin{pmatrix}
        a & b \\
        c & d
    \end{pmatrix} \in PSL(2, \mathbb{Z}) \middle| ~ c \equiv 0 \quad (\text{mod}~n)
    \right\}.
\end{align}
Moreover the subgroup $\bar{\Gamma}^0(n)$ which is written in Table \ref{tab:dualitygroups} is defined as follows:
\begin{align}
    \bar{\Gamma}^0(n) \equiv \left\{\begin{pmatrix}
        a & b \\
        c & d
    \end{pmatrix} \in PSL(2, \mathbb{Z}) \middle| ~ b \equiv 0 \quad (\text{mod}~n)
    \right\}.
\end{align}

\paragraph{Case II. $x_{43} \neq 0, x_{44} = 0$} \mbox{}\\

Repeating the same calculation, we can obtain the explicit matrix of $X$:
\begin{align}
    X_2 = \frac{1}{\sqrt{2} \sqrt{a d - b c}}
    \begin{pmatrix}
    0 & c & 0 & 2 d \\
    - 2 b & 0 & - 2 a & 0 \\
    0 & a & 0 & - 2 b \\
    d & 0 & c & 0\\
    \end{pmatrix}.
\end{align}
Since $X$ is the matrix of the symplectic transformation $Sp(4, \mathbb{Z})$ and these elements are integers, we must consider the following conditions for $a, b, c$ and $d$:
\begin{align}
    \begin{aligned}
        a'' &\equiv \frac{1}{2} \frac{\sqrt{2} a}{\sqrt{a d - b c}}, & b'' &\equiv \frac{\sqrt{2} b}{\sqrt{a d - b c}}, \\
        c'' &\equiv \frac{1}{2} \frac{\sqrt{2} c}{\sqrt{a d - b c}}, & d'' &\equiv \frac{1}{2} \frac{\sqrt{2} d}{\sqrt{a d - b c}},
    \end{aligned}
    \qquad  (a'', b'', c'', d'' \in \mathbb{Z}),
\end{align}
where these definitions lead $2 a'' d'' - b'' c'' = 1$.
By using $a'', b'', c''$ and $d''$, we can describe the symplectic modular transformation as the restricted linear fractional transformation:
\begin{align}
    X_2 = \begin{pmatrix}
    0 & c'' & 0 & 2 d'' \\
    - b ''& 0 & - 2 a'' & 0 \\
    0 & a'' & 0 & - b'' \\
    d'' & 0 & c'' & 0\\
    \end{pmatrix}.
\end{align}

Now we find that $Sp(4, \mathbb{Z})$ basis transformation can generate the transformation regarding the subgroup of $PSL(2, \mathbb{Z})$.
Here, we consider an element of the symplectic group $Sp(4, \mathbb{Z})$ as follows:
\begin{align}
    S_{\rm SD} = \begin{pmatrix}
        0 & 1 & 0 & 0 \\
        1 & 0 & 0 & 0 \\
        0 & 0 & 0 & 1 \\
        0 & 0 & 1 & 0 \\
    \end{pmatrix}.
\end{align}
Then it is found that this element can bridge between $X_1$ and $X_2$ as the following relation:
\begin{align}
    S_{\rm SD} x_1 = x_2, \qquad \left( x_1 \in X_1, ~ x_2 \in X_2 \right),
    \label{eq:bridge}
\end{align}
where these elements have a property of non-Abelian $S_{\rm SD} x_1 \neq x_1 S_{\rm SD}$.
Therefore, considering that the transformation of the dual cohomology basis is $Sp(4, \mathbb{Z})$, the modular transformation group of the complex-structure modulus can be denoted by $\bar{\Gamma}_0 (2)$ and $\mathbb{Z}_2$ regarding $S_{\rm SD}$ as follows:
\begin{align}
    \bar{\Gamma}_0 (2) \rtimes \mathbb{Z}_2.
\end{align}

Next, we focus on the transformation of $S_{\rm SD}$ as the linear fractional transformation.
Taking account of a transformation $S_{(2)} \equiv \frac{i}{\sqrt{2} U} S_{\rm SD}$ for the period vector, which includes the K\"{a}hler transformation, we find that this is associated with a modular transformation $U \rightarrow - \frac{1}{2 U}$ from the following calculation:
\begin{align}
    \begin{aligned}
        S_{(2)} \Pi' = \frac{i}{\sqrt{2} U}
        \begin{pmatrix}
            0 & 1 & 0 & 0 \\
            1 & 0 & 0 & 0 \\
            0 & 0 & 0 & 1 \\
            0 & 0 & 1 & 0 \\
        \end{pmatrix}
        \begin{pmatrix}
            1 \\
            -i\sqrt{2} U \\
            U \\
            \frac{i}{\sqrt{2}}
        \end{pmatrix}
        = \begin{pmatrix}
            1 \\
            \frac{i}{\sqrt{2} U} \\
            - \frac{1}{2 U} \\
            \frac{i}{\sqrt{2}}
        \end{pmatrix}.
    \end{aligned}
\end{align}
This transformation $U \rightarrow - \frac{1}{2 U}$ corresponds to the action of exchanging the inside and the outside of a circle of radius $\frac{1}{\sqrt{2}}$ and the circle has a fixed point $U = \frac{1}{2} + \frac{i}{2}$ of $\bar{\Gamma}_0 (2)$ in the circumference.
In other words, for $\bar{\Gamma}_0 (2)$, this can be understood as a natural generalization of the $S$-transformation of $PSL(2,Z)$.
Here, we call it ``Scaling duality'' \cite{Ishiguro:2023flux}.
For these results, the basis transformation regarding $Sp(4, \mathbb{Z})$ can be expressed naturally by the subgroup of $PSL(2, \mathbb{Z})$ and the generalized $S$-transformation.

\subsection{Conditions for the construction of the period vector}
\label{sec:conditions}

In this section, we discuss the general conditions necessary for the construction of period vectors, in order to extend the examples discussed in section \ref{sec:period_geometry} and section \ref{sec:period_modular} to the other orbifolds.
First, the conditions for constructing the third-cohomology basis of the toroidal orbifolds are summarized as follows:
\begin{itemize}
    \item The cycles of the toroidal orbifolds are composed of linear combinations of cycles on $T^6$, which have integer coefficients, because the toroidal orbifold is the subspace of $T^6$.

    \item Each cycle on the toroidal orbifold must be invariant under the Coxeter element $Q$.

    \item The third-cohomology basis of the toroidal orbifold has the symplectic structure as Eq. \eqref{eq:gneral_intersection}.
\end{itemize}
Although these conditions restrict the choice of the cycles regarding the toroidal orbifold, we can consider various linear combinations of the cycles of $T^6$, which satisfy the invariance of $Q$ and the symplectic structure.
Then, we choose the cycles which have the structure of bridging two symplectic transformation matrices by $S$-transformation or the Scaling duality as Eq. (\ref{eq:bridge}).
By considering this condition, we can discuss the generalization of the $S$-transformation naturally and understand the symplectic structure of the period vector obviously
\footnote{Elements of set generated by the symplectic transformation of the period vector or the modular transformation of the complex-structure modulus are also candidates of the period vector.}. 

The explicit form of the period vector can be chosen by the assumption that the period vector has a structure of the Scaling duality.
However, the period vector can not be fixed because we can take different integer values as the coefficient of overall of the cycles regarding the third-cohomology basis.
In addition, each period vector which has the different coefficient generates a different subgroup of the modular transformation.
This indicates that the symmetry emerging in the low-energy effective theory depends on the specific choice of coefficients for the overall of the cycles. Hence, the absence of a guiding principle for selecting these coefficients is a problem.

To resolve this problem, we discuss the consistency between the modular symmetry of the period vector and the duality symmetries.
Table \ref{tab:dualitygroups} shows the duality symmetries of the toroidal orbifolds which have a complex-structure modulus for the untwisted sector.
Here, our research can identify other duality symmetries that are distinct from those shown in previous research \cite{Bailin_1994}.
We discuss the details of the duality symmetries derived from the structure of the winding and the momentum numbers regarding the mass spectrum of the type IIB closed string in Appendix \ref{sec:duality}.
\begin{table}[H]
    \centering
    \begin{tabular}{|c|c|c|}
        \hline
        $\mathbb{Z}_N$, $\mathbb{Z}_N \times \mathbb{Z}_M$ & Lattice & Duality symmetries of $U$ \\ \hline
        $\mathbb{Z}_4$ & $SU(4)^2$ & $PSL(2, \mathbb{Z})$ \\
        $\mathbb{Z}_4$ & $SU(2) \times SU(4) \times SO(5)$ & $\bar{\Gamma}_0(2)$ \\
        $\mathbb{Z}_4$ & $SU(2)^2 \times SO(5)^2$ & $PSL(2, \mathbb{Z})$ \\
        $\mathbb{Z}_{6-\rm \greekii}$ & $SU(2) \times SU(6)$ & $\bar{\Gamma}_0(3)$ \\
        $\mathbb{Z}_{6-\rm \greekii}$ & $SU(3) \times SO(8)$ & $\bar{\Gamma}^0(3)$ \\
        $\mathbb{Z}_{6-\rm \greekii}$ & $SU(2)^2 \times SU(3)^2$ & $PSL(2, \mathbb{Z})$ \\
        $\mathbb{Z}_{6-\rm \greekii}$ & $SU(2)^2 \times SU(3) \times G_2$ & $PSL(2, \mathbb{Z})$ \\
        $\mathbb{Z}_{8-\rm \greekii}$ & $SU(2) \times SO(10)$ & $\bar{\Gamma}_0(2)$ \\
        $\mathbb{Z}_{8-\rm \greekii}$ & $SO(4) \times SO(9)$ & $PSL(2, \mathbb{Z})$ \\
        $\mathbb{Z}_{12-\rm \greekii}$ & $SO(4) \times F_4$ & $PSL(2, \mathbb{Z})$ \\
        $\mathbb{Z}_2 \times \mathbb{Z}_4$ & $SU(2)^2 \times SO(5)^2$ & $PSL(2, \mathbb{Z})$ \\
        $\mathbb{Z}_2 \times \mathbb{Z}_6$ & $SU(2)^2 \times SU(3) \times G_2$ & $PSL(2, \mathbb{Z})$ \\ \hline
    \end{tabular}
    \caption{The duality symmetries regarding the complex-structure modulus on $T^6/\mathbb{Z}_N$ and $T^6/(\mathbb{Z}_N \times \mathbb{Z}_M)$ orbifold.
    In the case of $SU(3) \times SO(8)$ loot lattice, the complex-structure modulus is defined as $U' \equiv U + 2$.}
    \label{tab:dualitygroups}
\end{table}
Taking account of the consistency, we must revisit definitions of the intersection number and the expansion of the holomorphic three-form.
If the above duality symmetries can be realized by the basis transformation of the period vector, it is necessary to discuss the way of constructing the third-cohomology basis and the holomorphic three-form as Eqs. \eqref{eq:integralsofsinglets}, \eqref{eq:gneral_intersection} and \eqref{eq:expansion}, which is not the usual normalization as $X^I = \int_{A^I} \Omega,~F_I =\int_{B_I} \Omega$.
Finally, by assuming these conditions, we can uniquely determine the period vector that satisfies the duality symmetry in the low-energy effective theory of string theory.
The additional conditions taken into account are shown as follows:
\begin{itemize}
    \item The period vector has a structure where the Scaling duality bridges the symplectic transformation matrices of the period vector.
    In other words, these actions are realized as an outer automorphism of the symplectic transformation group $Sp(4, \mathbb{Z})$.

    \item The modular transformation of the period vector is consistent with the duality symmetries of the complex-structure modulus.
\end{itemize}
For the other orbifolds, we summarize the results of these calculations in Appendix \ref{sec:orbifolds}.

\section{Flux landscape of type IIB string theory}
\label{sec:landscape}

\subsection{Effective action and three-form fluxes $G_3$}
\label{sec:effective_action}

In this section, we discuss the construction of effective potentials induced by three-form fluxes by using the third-cohomology basis and the holomorphic three-form which are constructed in section \ref{sec:period_geometry} and section \ref{sec:period_modular}.
Here, we define the three-form as $G_3 \equiv F_3 - S H_3$ where $F_3$ denotes the Ramond-Ramond three-form, $H_3$ denotes the three-form in the Neveu-Scwarz sector, and $S \equiv C_0 + i~e^{-\phi}$ is the axio-dilaton.
In the effective Lagrangian of type IIB supergravity, the K\"{a}hler potential $K$ and the Gukov-Vafa-Witten superpotential $W$ \cite{Gukov:1999ya} are denoted by
\begin{align}
    K &= - \log\left(-i \left( S - \bar{S} \right)\right) - \log\left(-i \int{\Omega \wedge \bar{\Omega}}\right), \\
    W &= \int{G_3 \wedge \Omega},
    \label{eq:superpotential}
\end{align}
and the no-scale type scalar potential in the unit of $M_{\mathrm{Pl}}=1$ is also denoted by 
\begin{align}
    V = e^K \left(K^{I \bar{J}} D_I W D_{\bar{J}} \bar{W} \right),
\end{align}
where the labels $I, J$ include the axio-dilaton and the complex-structure modulus.
$K^{IJ}$ is the inverse of the K\"{a}hler metric $K_{I \bar{J}} \equiv \partial_I \partial_{\bar{J}} K$ and $D_I W \equiv \partial W + K_I W$ is the covariant derivative of $W$.

In this study, we discuss the quantization of the three-form fluxes of Eq. \eqref{eq:superpotential} on the cycles of $T^6$ as follows:
\begin{equation}
    \begin{alignedat}{5}
        \int_{T^6/\mathbb{Z}_4}{(F_3, H_3) \wedge \alpha_i} &\equiv (-b_i, -d_i), \quad    &  \int_{T^6/\mathbb{Z}_4}{(F_3, H_3) \wedge \beta^i} &\equiv (a^i, c^i),  \\
        \int_{T^6/\mathbb{Z}_4}{(F_3, H_3) \wedge \gamma_i} &\equiv (-f_i, -h_i), \quad    &  \int_{T^6/\mathbb{Z}_4}{(F_3, H_3) \wedge \delta^i} &\equiv (e^i, g^i).
    \end{alignedat}
    \label{eq:realfluxquanta}
\end{equation}
Here, we consider the quantization which has no contribution of the twisted sector associated with the complex structure moduli.
By using these definitions, the three-form fluxes on the cycles which are invariant under the Coxeter element $Q$ can be denoted by
\begin{align}
    \begin{alignedat}{2}
        \frac{1}{4} \int_{T^6/\mathbb{Z}_4} (F_3, H_3) \wedge \mathbf{1}_{A^0} &= \frac{1}{8} \int_{T^6/\mathbb{Z}_4} (F_3, H_3) \wedge \left[ \sum \Gamma(\alpha_1) \right] \\
        &= \frac{1}{8} \int_{T^6/\mathbb{Z}_4} (F_3, H_3) \wedge \left[ Q (\alpha_1) + Q^2 (\alpha_1) + Q^3 (\alpha_1) + Q^4 (\alpha_1) \right] \\
        &= \frac{1}{2} (- b_1, - d_1), \\
        \frac{1}{8} \int_{T^6/\mathbb{Z}_4} (F_3, H_3) \wedge \mathbf{1}_{A^1} &= \frac{1}{2}(a^1, c^1), \\
        \frac{1}{4} \int_{T^6/\mathbb{Z}_4} (F_3, H_3) \wedge \mathbf{1}_{B_0} &= (e^2, g^2), \\
        \frac{1}{8} \int_{T^6/\mathbb{Z}_4} (F_3, H_3) \wedge \mathbf{1}_{B_1} &= \frac{1}{2} (- b_0, - d_0).
    \end{alignedat}
    \label{eq:integralsforfluxquanta-Z4}
\end{align}
The overall factors $1/4$ and $1/8$ are terms to offset overcounting of the intersection numbers Eq. \eqref{eq:integralsofsinglets} as the same discussion as in Eq. \eqref{eq:generalizedcoordinates} and Eq. \eqref{eq:expansion}.
In the second line of Eq. \eqref{eq:integralsforfluxquanta-Z4}, the number of the elements of $\Gamma(\alpha_0)$ is found by the length of orbit in Eq. \eqref{eq:lengthoforbit}.
In addition, we know a relation $Q(dx^1 \wedge dx^2 \wedge dx^3 \wedge dx^4 \wedge dx^5 \wedge dx^6) = dx^1 \wedge dx^2 \wedge dx^3 \wedge dx^4 \wedge dx^5 \wedge dx^6$ and this result derives $\int_{T^6/\mathbb{Z}_4} (F_3, H_3) \wedge Q^n (\alpha_1) = \int_{T^6/\mathbb{Z}_4} (F_3, H_3) \wedge \alpha_1$.
Thus, the three-form fluxes on the invariant cycles under $Q$ are denoted by the flux quanta of $T^6$.
Taking account of the quantization of these three-form fluxes, it is necessary to choose specific constant multiples for the flux quanta as follows:
\begin{align}
        (b_1, d_1) \in 4 \mathbb{Z},\quad (a^1, c^1) \in 4 \mathbb{Z},\quad (e^2, g^2) \in 2 \mathbb{Z},\quad  (b_0, d_0) \in 4 \mathbb{Z}.
    \label{eq:fluxquantizationcondition-Z4}
\end{align}
Note that, for simplicity, we consider the additional even multiples for the flux quanta in order to avoid the exotic $O3$-planes \cite{Frey:2002hf, Kachru:2002he, Cascales:2003zp, Font:2004cy, Ishiguro:2023flux}.

By using these results, we can construct the expansion of $G_3$ as 
\begin{align}
   \begin{aligned}
        G_3 =~& (e^2 - S g^2) \mathbf{1}_{A^0} - \frac{1}{2} (b_0 - S d_0) \mathbf{1}_{A^1} + \frac{1}{2} (b_1 - S d_1) \mathbf{1}_{B_0} - \frac{1}{2} (a^1 - S c^1) \mathbf{1}_{B_1} \\
        =~& (e^2 - S g^2) \left( \alpha_0 - \gamma_2 - 2 \gamma_3 - \delta^5 - \delta^6 \right) - \frac{1}{2} (b_0 - S d_0) \left( -2 \beta^0 - \gamma_5 + \gamma_6 - \delta^3 \right) \\
        &+ \frac{1}{2} (b_1 - S d_1) \left( 2 \beta^1 + \gamma_5 - \gamma_6 - 2 \delta^2 - \delta^3 \right) - (a^1 - S c^1) \left( \alpha_0 - \alpha_1 -\delta^5 + \delta^6 \right).
   \end{aligned}
\end{align}
Then, we can describe $N_{\rm flux}$ as follows:
\begin{align}
    \begin{aligned}
         N_{\rm flux} &= \int H_3 \wedge F_3 = 2( a^1 d_0 -c^1 b_0 -g^2 b_1 + e^2 d_1 ) \in 16 \mathbb{Z}.
    \end{aligned}
\end{align}
Since we can obtain the three-form fluxes and the holomorphic three-form, the superpotential can be represented by them.
The general form of the superpotential is expressed as follows:
\begin{align}
    W = A + B S + U \left[ C + D S \right],
    \label{eq:WGeneral}
\end{align}
and $A, B, C$ and $D$ in the case of $T^6/\mathbb{Z}_4$ orientifold with $SU(2) \times SU(4) \times SO(5)$ root lattice can be found by the calculation of Eq. (\ref{eq:superpotential}) as follows:
\begin{equation}
    \begin{alignedat}{5}
        A &= - \frac{1}{\sqrt{2}} (b_1 - i b_0) , \quad & C &= - \sqrt{2} ( e^2 + i a^1 ), \\
        B &= \frac{1}{\sqrt{2}} (d_1 - i d_0) , \quad & D &= \sqrt{2} ( g^2 + i c^1 ).
    \end{alignedat}
    \label{eq:ABCDinrealfluxes}
\end{equation}

Taking account of the landscape of type IIB supergravity, the stationary point $\partial_I V = 0$ leads to the VEVs of the axio-dilaton and the complex-structure modulus. 
It is possible to obtain them by solving the following F-term equations:
\begin{equation}
    \left\{
    \begin{alignedat}{3}
        D_U W &= \partial_U W + K_U W &= 0\\
        D_S W & = \partial_S W + K_S W &= 0\\
        W &= 0 
    \end{alignedat}
    \right.
    \quad \Rightarrow \quad
    \left\{
    \begin{alignedat}{2}
        C + D S &= 0\\
        B + D U &= 0\\
        A + C U &= 0
    \end{alignedat}\right.
    , \label{eq:ftermeq}
\end{equation}
which is an overdetermined system. 
Therefore, we can derive the flux landscape of SUSY Minkowski solutions regarding the axio-dilaton and the complex-structure moduli as follows:
\begin{align}
    \langle S \rangle = - \frac{C}{D}, \quad \langle U \rangle = - \frac{B}{D}, \quad A D - B C = 0.
    \label{eq:ftermsolution}
\end{align}

\subsection{Flux transformations for symmetries in the F-term equations}
\label{sec:flux_transformations}

For the modular symmetries of the complex-structure modulus in the type IIB flux landscape, it is not enough to discuss the transformation of the period vector because the three-form fluxes on the cycles of orientifold break the modular symmetry.
Focusing on the F-term equation regarding the complex-structure modulus, it can be found from the following equation
\begin{align}
    \begin{aligned}
        D_U W &= - \frac{1}{U - \overline{U}} \left( A + B S + \overline{U} (C + D S)\right) \\
        &=- \frac{1}{U - \overline{U}}
        \begin{pmatrix}
            1 & \overline{U}
        \end{pmatrix}
        \begin{pmatrix}
            A & B \\
            C & D \\
        \end{pmatrix}
        \begin{pmatrix}
            1 \\
            S
        \end{pmatrix}
        = 0.
    \end{aligned}
\end{align}
Then, under the liner fractional transformation of the complex-structure modulus as
\begin{align}
    \begin{aligned}
        U \rightarrow U' = \frac{a U + b}{c U + d},
    \end{aligned}
    \qquad ( a, b , c, d \in \mathbb{Z}, \quad a d - b c = 1 ),
\end{align}
we require the following flux transformations for $A, B, C$ and $D$,
\begin{align}
    \begin{pmatrix}
        A' & B' \\
        C' & D'
    \end{pmatrix}
    = 
    \begin{pmatrix}
        a & -b\\
        -c & d
    \end{pmatrix}
    \begin{pmatrix}
        A & B \\
        C & D
    \end{pmatrix}.
    \label{eq:fluxesABCDtransformation-sl2z-cs}
\end{align}
By this requirement for the three-form fluxes, we can obtain the covariant F-term equations under the modular transformation and classify the modular symmetries derived from the geometrical structures of the various orientifolds in the type IIB flux landscape
\footnote{For the axio-dilaton $S$, we can obtain the transformations regarding the three-form fluxes by the same discussion.}. 

Let us summarize the flux transformations in terms of the real flux quanta (\ref{eq:realfluxquanta}). For the two generators $S, T$ of $SL(2, \mathbb{Z})_{S}$,
\begin{itemize}
    \item $S = 
        \begin{pmatrix}
            0 & -1 \\
            1 & 0
        \end{pmatrix} \in SL(2, \mathbb{Z})_S
    $\\
    \begin{equation}
        \begin{alignedat}{5}
            a^1 &\rightarrow - c^1, \quad & b_0 &\rightarrow - d_0, \\
            c^1 &\rightarrow a^1, \quad & b_1 &\rightarrow - d_1, \\
            e^2 &\rightarrow - g^2, \quad & d_0 &\rightarrow b_0, \\
            g^2 &\rightarrow e^2, \quad & d_1 &\rightarrow b_1. \\
        \end{alignedat}
    \end{equation}
\end{itemize}
\begin{itemize}
    \item $T^q = 
        \begin{pmatrix}
            1 & q \\
            0 & 1
        \end{pmatrix} \in SL(2, \mathbb{Z})_S
    $\\
    \begin{equation}
        \begin{alignedat}{5}
            a^1 &\rightarrow a^1 + q c^1, \quad & b_0 &\rightarrow b_0 + q d_0, \\
            c^1 &\rightarrow c^1, \quad & b_1 &\rightarrow b_1 + q d_1, \\
            e^2 &\rightarrow e^2 + q g^2, \quad & d_0 &\rightarrow d_0, \\
            g^2 &\rightarrow g^2, \quad & d_1 &\rightarrow d_1. \\
        \end{alignedat}
    \end{equation}
\end{itemize}
For those of $SL(2, \mathbb{Z})_U$, 
\begin{itemize}
    \item $S = 
        \begin{pmatrix}
            0 & -1 \\
            1 & 0
        \end{pmatrix} \in SL(2, \mathbb{Z})_U
    $\\
    \begin{equation}
        \begin{alignedat}{5}
            a^1 &\rightarrow \frac{b_0}{2}, \quad & b_0 &\rightarrow - 2 a^1, \\
            c^1 &\rightarrow \frac{d_0}{2}, \quad & b_1 &\rightarrow 2 e^2, \\
            e^2 &\rightarrow - \frac{b_1}{2}, \quad & d_0 &\rightarrow - 2 c^1, \\
            g^2 &\rightarrow - \frac{d_1}{2}, \quad & d_1 &\rightarrow 2 g^2. \\
        \end{alignedat}
        \label{eq:fluxtrf-S}
    \end{equation}
\end{itemize}
\begin{itemize}
    \item $T^q = 
        \begin{pmatrix}
            1 & q \\
            0 & 1
        \end{pmatrix} \in SL(2, \mathbb{Z})_U
    $\\
    \begin{equation}
        \begin{alignedat}{5}
            a^1 &\rightarrow a^1, \quad & b_0 &\rightarrow b_0 + 2 q a^1, \\
            c^1 &\rightarrow c^1, \quad & b_1 &\rightarrow b_1 - 2 q e^2, \\
            e^2 &\rightarrow e^2, \quad & d_0 &\rightarrow d_0 + 2 q c^1, \\
            g^2 &\rightarrow g^2, \quad & d_1 &\rightarrow d_1 - 2 q g^2. \\
        \end{alignedat}
        \label{eq:fluxtrf-T}
    \end{equation}
\end{itemize}
For the $V_1$-transformation of $\Gamma_0(2)_U$, 
\begin{itemize}
    \item $V_1 = 
        \begin{pmatrix}
            1 & 1 \\
            -2 & -1
        \end{pmatrix} \in \Gamma_0(2)_U
    $\\
    \begin{equation}
        \begin{alignedat}{5}
            a^1 &\rightarrow - a^1 - b_0, \quad & b_0 &\rightarrow b_0 + 2 a^1, \\
            c^1 &\rightarrow - c^1 - d_0, \quad & b_1 &\rightarrow b_1 - 2 e^2, \\
            e^2 &\rightarrow - e^2 + b_1, \quad & d_0 &\rightarrow d_0 + 2 c^1, \\
            g^2 &\rightarrow - g^2 + d_1, \quad & d_1 &\rightarrow d_1 - 2 g^2. \\
        \end{alignedat}
        \label{eq:fluxtrf-Sprime}
    \end{equation}
\end{itemize}

For the Scaling duality, the transformation regarding the complex-structure modulus is $S_{(2)} : U \rightarrow - \frac{1}{2U} \in SL(2, \mathbb{R})$. 
Note that it does not belong to $SL(2, \mathbb{Z})$ because these elements for the linear fractional transformation are $a, d = 0$ and $b=-\frac{1}{\sqrt{2}}, c = \sqrt{2}$.
In this research, since we consider the scalar potential which is invariant under the Scaling duality, the flux transformations include the coefficients of $\sqrt{2}$ and $\frac{1}{\sqrt{2}}$.
Under the $S_{(2)}$, the three-form fluxes transform as
\begin{itemize}
    \item $S_{(2)} = 
        \begin{pmatrix}
            0 & -\frac{1}{\sqrt{2}} \\
            \sqrt{2} & 0
        \end{pmatrix} \notin \Gamma_0(2)_U
    $\\
    \begin{equation}
        \begin{alignedat}{5}
            a^1 &\rightarrow \frac{b_0}{\sqrt{2}}, \quad & b_0 &\rightarrow - \sqrt{2} a^1, \\
            c^1 &\rightarrow \frac{d_0}{\sqrt{2}}, \quad & b_1 &\rightarrow \sqrt{2} e^2, \\
            e^2 &\rightarrow - \frac{b_1}{\sqrt{2}}, \quad & d_0 &\rightarrow - \sqrt{2} c^1, \\
            g^2 &\rightarrow - \frac{d_1}{\sqrt{2}}, \quad & d_1 &\rightarrow \sqrt{2} g^2. \\
        \end{alignedat}
        \label{eq:fluxtrf-Sthree}
    \end{equation}
\end{itemize}

\section{Conclusion}
\label{sec:conc}

In this paper, we investigated the construction of the period vector and the modular symmetries on various toroidal orbifolds.
Since the action of the low-energy effective theory associated with type IIB string theory is described by the period vector, the research of these symmetries corresponds to revealing the symmetries appearing in the low-energy effective theory.
In particular, it is known that the flux landscape of type IIB supergravity tends to be realized at fixed points in the modular transformation groups \cite{Ishiguro:2020tmo, Ishiguro:2023flux}.
It motivates us to construct the period vectors which are consistent with string theory and examine the modular symmetries emerging from the geometric structure of the toroidal orbifolds.

In section \ref{sec:period_geometry} and section \ref{sec:period_modular}, we constructed the period vector on $T^6/\mathbb{Z}_4$ orbifold with $SU(2) \times SU(4) \times SO(5)$ root lattice.
Although the explicit period vector can be obtained by using the traditional discussions regarding the holomorphic three-form and the third-cohomology basis, there is redundancy concerning the linear combinations of the cycles in the third-cohomology basis.
Then we partially eliminated this redundancy by choosing the cycles that have the structure of ``Scaling duality'' in section \ref{sec:trans_period} because we can discuss the natural generalization for $S$-transformation of $PSL(2, \mathbb{Z})$.
Note that there is still redundancy that some period vectors can be transformed into the desired period vector by the symplectic transformation group $Sp(4, \mathbb{Z})$.

In addition, we must consider the problem that the overall coefficients of the cycles are unfixed.
Since the overall factors are associated with the intersection numbers and the relative differences regarding the elements of the period vectors, the different overall factors generate different modular symmetries.
For this point, the factors of the cycles in $T^6/\mathbb{Z}_4$ orbifold can be determined by choosing the period vector that the modular symmetry is the same as the duality symmetries.
Here, we need to carefully discuss the definition of the coordinates of each cycle concerning the period vector in section \ref{sec:const_period}.
In terms of modular symmetry, this condition allows the realization of the low-energy effective theory that is consistent with string theory.
By using these conditions, the period vectors can be constructed on some toroidal orbifolds and we can conclude that the structure of ``Scaling duality'' and the consistency with duality symmetries is necessary to discuss the systematic classification of the modular symmetries regarding the period vectors.

Furthermore, by taking account of the cycles on the toroidal orbifolds, we defined the three-form fluxes $G_3$ and constructed the Gukov-Vafa-Witten superpotential.
Then it is found that the flux quanta take the specific constant multiples on the cycles of $T^6/\mathbb{Z}_4$ orientifold as same as the discussion in \cite{Ishiguro:2023flux}.
In addition, by revealing the explicit flux transformations for the modular transformation of the complex-structure modulus, we can realize the modular symmetries in the type IIB flux landscape.
In our future work, we will discuss the Dirac quantization of the three-form fluxes and the tadpole cancellation condition in detail because we only consider the brief discussion of the flux quanta to avoid the exotic $O3$-plane.

\acknowledgments

This work was supported in part by K2-SPRING program Grant number JPMJSP2136 (T. Kai) and JSPS KAKENHI Grant Numbers JP23H04512 (H.O) and JP23K03375 (T. Kobayashi).

\appendix

\section{The duality symmetry from the partition function}
\label{sec:duality}
In this Appendix, we review the duality symmetry derived from closed string partition functions based on \cite{Mayr:1993mq,Bailin_1994,Erler_1993} whose main subjects are the orbifold compactification of heterotic strings.
The partition function for the bosonic closed string part in type II can be obtained from ten-dimensional bosonic one of heterotic strings.

The starting point is the one-loop partition function for type IIB closed strings on the six-dimensional torus
\begin{equation}
    Z_{\mathrm{IIB}}(\tau,\bar{\tau},G,B)=\frac{1}{4\tau_2(\eta\bar{\eta})^{12}}\left|\vartheta_3^4-\vartheta_4^4-\vartheta_2^4\right|^{2}\sum_{P\in\Lambda}q^{\frac{1}{2}P_L^2}\bar{q}^{\frac{1}{2}P_R^2},
\end{equation}
where $\tau=\tau_1+i\tau_2$ is the modulus of the world–sheet torus and $q=e^{2\pi i\tau}$. The $\tau$-dependent functions $\eta(\tau)$ and $\vartheta_i(\tau)$ are the Dedekind eta function and the $i$-th elliptic theta function, respectively. The summation of $P$ over the lattice $\Lambda$ comes from the internal directions and the element $P$ of the lattice is expressed as
\begin{subequations}
    \begin{align}\label{momentum}
       & P_{L}=\frac{p}{2}+(G-B)w, \\
       & P_{R}=\frac{p}{2}-(G+B)w,
    \end{align}
\end{subequations}
where $w$ and $p$ are integer six vectors taking values on the lattice $\Lambda$ and its dual $\Lambda^{*}$, corresponding to the winding numbers and the Kaluza-Klein momenta, and the background fields $G$ and $B$ are a metric of the torus and an anti-symmetric tensor, called the moduli of the torus. The lattice $\Lambda$, called the Narain lattice, must be even and self-dual by the modular invariance of the partition function.

From now on, we consider the $T^6/\mathbb{Z}_{N}$ orbifolds for simplicity, but the following discussion can easily be generalized to the $\mathbb{Z}_{M}\times\mathbb{Z}_{N}$ orbifolds. In general, the partition function for the orbifolds can be written as the sum of all of the sectors projected by $\mathbb{Z}_{N}$ twists $\theta$:
\begin{align}
    Z_{\mathrm{orb}}(\tau,\bar{\tau},G,B)=\frac{1}{N}\sum_{k,l=0}^{N-1}Z[\theta^k,\theta^l],
\end{align}
where the blocks $Z[1,\theta^l]$ are called the untwisted sectors while the other blocks correspond to the contributions from the twisted string states, which can be obtained by the modular transformations of the untwisted sectors. An important fact is that if $l$ is a divisor of $N$, $Z[1,\theta^l]$ has the dependence of the moduli $G,B$ which comes from the summation of the momentum lattice, and then the twisted sector obtained from that untwisted sector also has the moduli dependence. In these sectors, some supersymmetries are preserved since $\theta^l$ leaves some fixed directions when $l$ is a divisor of $N$. In order to study the duality symmetry, it is enough to focus on those moduli-dependent parts of the partition functions.

The orbifold twists $\theta$ are realized by Coxeter elements $Q$ that specify the orbifold point group. In order to ensure that the point group is an automorphism of the lattice, the background $G,B$ must satisfy  
\begin{align}
    Q^{t}GQ=G,~~~Q^{t}BQ=B.
\end{align}
Solving these conditions, one can find that the moduli are (partially) fixed. The moduli-dependent part in the untwisted sectors can be expressed as
\begin{align}
    Z_{(1,\theta^{l})}(\tau,\bar{\tau},G,B)=\sum_{P\in\Lambda^{\perp}}q^{\frac{1}{2}P_L^tG^{-1}P_L}\bar{q}^{\frac{1}{2}P_R^tG^{-1}P_R},
\end{align}
where $\Lambda^{\perp}$ is the sub-lattice of $\Lambda$ invariant under $Q^l$ and we assume $l$ to be a divisor of $N$. From the invariance under orbifold twists, the windings $w$ and momenta $p$ must satisfy
\begin{align}
    Q^lw=w,~~~((Q^t)^{-1})^lp=p,
\end{align}
which lead to the fixed directions and the duality groups of the orbifold theories.

As a concrete example, we take the $T^6/\mathbb{Z}_4$ orbifold with the $SU(2) \times SU(4) \times SO(5)$ root lattice, as discussed in sections \ref{sec:period_geometry} and \ref{sec:period_modular}. In this case, the moduli-dependent parts in the untwisted sectors are $Z_{(1,1)},Z_{(1,\theta^{2})}$ and in the twisted sectors are $Z_{(\theta^{2},1)},Z_{(\theta^{2},\theta^{2})}$, which can be obtained by a $S$- and $ST$-transformation from $Z_{(1,\theta^{2})}$, respectively. We can focus on $Z_{(1,\theta^{2})}$ to identify the duality group of this theory. 

Conditions $Q^2w=w$ and $((Q^t)^{-1})^2p=p$ give fixed directions (\ref{w&p}), where $Q$ is given by (\ref{eq:matrixQ}). The invariant lattice $\Lambda^{\perp}$ under $Q^2$ gives the two-dimensional sub-torus with the metric $G_{\perp}$ and $B$-field $B_{\perp}$ defined by $w^tGw=(n_1~n_2)G_{\perp}(n_1~n_2)^t$ and $p^tBp=(m_1~m_2)B_{\perp}(m_1~m_2)^t$. We introduce the K\"ahler and complex structure moduli $U$ and $T$ defined as
\begin{align}
    T=2(B_{\perp12}+i\sqrt{\mathrm{det}G_{\perp}}),~~~U=\frac{1}{G_{\perp11}}(G_{\perp12}+i\sqrt{\mathrm{det}G_{\perp}}).
\end{align}
Using these moduli, we can write down $Z_{(1,\theta^{2})}$ as
\begin{align}
    Z_{(1,\theta^{2})}(\tau,\bar{\tau},T,U)=\sum_{m_1,m_2,n_1,n_2\in\mathbb{Z}}e^{2\pi i\tau_1(2m_1n_1+m_2n_2)}e^{-\frac{\pi\tau_{2}}{T_2U_2}|TUn_2+Tn_1-2Um_1+m_2|^2},
\end{align}
where $U=U_1+iU_2$ and $T=T_1+iT_2$. From this partition function, we can read off the mass spectrum of strings for the two-dimensional sub-torus
\begin{align}
    m_{\perp}^2=\sum_{m_1,m_2,n_1,n_2\in\mathbb{Z}}\frac{1}{T_2U_2}|TUn_2+Tn_1-2Um_1+m_2|^2,
\end{align}
which is invariant under the following transformations on $T$ and $U$:
\begin{subequations}
    \begin{align}
        &T\to T+2,~~~T\to\frac{T}{T+1},\\
        &U\to U+1,~~~U\to-\frac{U}{2U-1}.
    \end{align}
\end{subequations}
Therefore, we conclude that the duality group for this orbifold is given by $\Gamma^0(2)_T\times\Gamma_0(2)_U$.

The duality symmetries in the other cases can also be studied in the same way.

\section{Various orbifolds}
\label{sec:orbifolds}

In this appendix, we show the period vectors and three-from fluxes in various orbifold compactifications with $ h^{2,1}_{\text{untw.}} =1$.

\subsection{$T^6/\mathbb{Z}_4$ orbifold with $SU(4)^2$ root lattice}

Considering $T^6/\mathbb{Z}_4$ orbifold, it is required that the six-dimensional metric is invariant under the action of $\Gamma = \mathbb{Z}_4$.
By using the Coxeter element $Q$ corresponding to $SU(4)^2$ root lattice, we can define $\mathbb{Z}_4$ basis transformation for the metric.
The action of $Q$ on $SU(4)^2$ root lattice is defined as
\begin{equation}
    \begin{aligned}
     Q(e_1) &= e_2, \quad Q(e_2) &= e_3, \quad Q(e_3) &= - e_1 - e_2 - e_3, \\
     Q(e_4) &= e_5, \quad Q(e_5) &= e_6, \quad Q(e_6) &= - e_4 - e_5 - e_6.
    \end{aligned}
\end{equation}
The matrix representation of $Q$ is defined by $Q(e_i) = e_j Q_{ji}$ and the explicit matrix is shown below:
\begin{align}
    Q = \begin{pmatrix}
        0 & 0 & -1 & 0 & 0 & 0 \\
        1 & 0 & -1 & 0 & 0 & 0 \\
        0 & 1 & -1 & 0 & 0 & 0 \\
        0 & 0 & 0 & 0 & 0 & -1 \\
        0 & 0 & 0 & 1 & 0 & -1 \\
        0 & 0 & 0 & 0 & 1 & -1 \\
    \end{pmatrix},
\end{align}
where $Q^4 = 1$.

To construct the cycles on the orbifold, we apply this action $\sum \Gamma$ to 20 real three-forms.
By choosing only the linearly independent cycles from the obtained invariant cycles, the following four cycles are derived;
\begin{align}
    \begin{aligned}
    \mathbf{1}_{A^0} &\equiv \sum \Gamma(\alpha_3), \\
    \mathbf{1}_{A^1} &\equiv \sum \Gamma(\beta^1), \\
    \mathbf{1}_{B_0} &\equiv - \sum \Gamma(\alpha_2), \\
    \mathbf{1}_{B_1} &\equiv \sum \Gamma(\beta^2) .
    \end{aligned}
\end{align}
Here, we also consider the sum of orbits associated with the Coxeter element and the lengths of four orbits are
\begin{align}
    |\Gamma(\alpha_2)| = 4, \quad |\Gamma(\alpha_3)| = 4, \quad |\Gamma(\beta^1)| = 4, \quad |\Gamma(\beta^2)| = 4.
\end{align}
In terms of 20 real three-forms, these cycles can be expressed as follows:
\begin{align}
    \begin{aligned}
    \mathbf{1}_{A^0} &= \alpha_0 + \alpha_3 + \beta^2 - \gamma_2 - \gamma_3 - \delta^5, \\
    \mathbf{1}_{A^1} &= \alpha_2 - \beta^0 + \beta^1 - \gamma_4 - \delta^2 - \delta^3, \\
    \mathbf{1}_{B_0} &= \beta^0 + \beta^1 + \beta^3 +4 \gamma_5 - \gamma_6 - \delta^2, \\
    \mathbf{1}_{B_1} &= \alpha_0 - \alpha_1 -\alpha_3 + \gamma_3 - \delta^4 + \delta^6.
    \end{aligned}
\end{align}
On $T^6/\mathbb{Z}_4$, these intersections of the dual cycles in regards to the orbifold satisfy the following relations;
\begin{align}
    \int_{T^6/\mathbb{Z}_4} \mathbf{1}_{A^0} \wedge \mathbf{1}_{B_0} = 4, \quad \int_{T^6/\mathbb{Z}_4} \mathbf{1}_{A^1} \wedge \mathbf{1}_{B_1} = 4,
\end{align}
where the other intersection numbers are 0.

Considering the discussion in section \ref{sec:holomorphic_three-form}, we can construct the complex one-form as follows:
\begin{align}
    \begin{alignedat}{2}
    dz^1 &= \frac{1}{\sqrt{2}} \left( dx^1 + i ~ dx^2 - dx^3 \right), \\
    dz^2 &= \frac{1}{\sqrt{2}} \left( dx^4 + i ~ dx^5 - dx^6 \right), \\
    dz^3 &= \frac{1}{2} \left[ dx^1 - dx^2 + dx^3 + U \left( dx^4 - dx^5 + dx^6 \right) \right].
    \end{alignedat}
\end{align}
Consequently, by using these complex one-forms, the holomorphic three-form is defined as $\Omega = dz^1 \wedge dz^2 \wedge dz^3$ where $\int_{T^6/\mathbb{Z}_4} \Omega \wedge \bar{\Omega} = 2 i \text{Im} U$.
Moreover, the period vector is denoted by
\begin{align}
    \begin{aligned}
        \Pi &\equiv \begin{pmatrix}
               X^0 \\
               X^1 \\
               F_0 \\
               F_1
               \end{pmatrix}
            &= \begin{pmatrix}
               \frac{1}{2} \int_{T^6/\mathbb{Z}_4} \Omega \wedge \mathbf{1}_{B_0} \\
               \frac{1}{2} \int_{T^6/\mathbb{Z}_4} \Omega \wedge \mathbf{1}_{B_1} \\
               \frac{1}{2} \int_{T^6/\mathbb{Z}_4} \Omega \wedge \mathbf{1}_{A^0} \\
               \frac{1}{2} \int_{T^6/\mathbb{Z}_4} \Omega \wedge \mathbf{1}_{A^1}
               \end{pmatrix} 
            &= \begin{pmatrix}
                \frac{1 - i}{2} \\
                - \frac{1 + i}{2} U \\
                \frac{1 - i}{2} U \\
                \frac{1 + i}{2}
               \end{pmatrix},
    \end{aligned}
\end{align}
which satisfies $\Pi^{\dagger} \Sigma \Pi = 2 i \text{Im} U$.
Since the overall factor can be absorbed by K\"{a}ler transformation, we consider a modified period vector $\Pi' = \left(1, -i U, U', i \right)^t$ in the discussion of modular transformation.

Here, we consider the quantization which has no contribution of the twisted sector regarding $h^{2,1}$.
By using these definitions, the three-form fluxes on the cycles which are invariant under the Coxeter element $Q$ can be denoted by
\begin{align}
    \begin{alignedat}{2}
        \frac{1}{4} \int_{T^6/\mathbb{Z}_4} (F_3, H_3) \wedge \mathbf{1}_{A^0} &= (- b_3, - d_3), \\
        \frac{1}{4} \int_{T^6/\mathbb{Z}_4} (F_3, H_3) \wedge \mathbf{1}_{A^1} &= (a^1, c^1), \\
        \frac{1}{4} \int_{T^6/\mathbb{Z}_4} (F_3, H_3) \wedge \mathbf{1}_{B_0} &= - (- b_2, - d_2), \\
        \frac{1}{4} \int_{T^6/\mathbb{Z}_4} (F_3, H_3) \wedge \mathbf{1}_{B_1} &= (a^2, c^2).
    \end{alignedat}
\end{align}
Taking account of the quantization of these three-form fluxes, it is necessary to choose specific constant multiples for the flux quanta as follows:
\begin{align}
        (a^1, c^1) \in 2 \mathbb{Z},\quad (a^2, c^2) \in 2 \mathbb{Z},\quad (b_2, d_2) \in 2 \mathbb{Z},\quad (b_2, d_2) \in 2 \mathbb{Z}.
\end{align}
By using these results, we can construct the expansion of $G_3$ as 
\begin{align}
   \begin{aligned}
        G_3 =~& (b_2 - S d_2) \mathbf{1}_{A^0} - (a^2 - S c^2) \mathbf{1}_{A^1} + (b_3 - S d_3) \mathbf{1}_{B_0} - (a^1 - S c^1) \mathbf{1}_{B_1}.
   \end{aligned}
\end{align}
Then, we can describe $N_{\rm flux}$ as follows:
\begin{align}
    \begin{aligned}
         N_{\rm flux} &= \int H_3 \wedge F_3 = 4 (-a^2 c^1 + a^1 c^2 + b_3 d_2 - b_2 d_3) \in 16 \mathbb{Z}.
    \end{aligned}
\end{align}
Since we can obtain the three-form fluxes and the holomorphic three-form, the superpotential can be denoted by them.
$A, B, C$ and $D$ in the case of $T^6/\mathbb{Z}_4$ orientifold with $SU(4)^2$ root lattice can be found as follows:
\begin{equation}
    \begin{alignedat}{5}
        A &= (1 + i) a^2 - (1 - i) b_3, \quad & C &= - (1 + i) a^1 - (1 - i) b_2, \\
        B &= - (1 + i) c^2 + (1 - i) d_3, \quad & D &= (1 + i) c^1 + (1 - i) d_2.
    \end{alignedat}
\end{equation}

\subsection{$T^6/\mathbb{Z}_4$ orbifold with $SU(2)^2 \times SO(5)^2$ root lattice}

Considering $T^6/\mathbb{Z}_4$ orbifold, it is required that the six-dimensional metric is invariant under the action of $\Gamma = \mathbb{Z}_4$.
By using the Coxeter element $Q$ corresponding to $SU(2)^2 \times SO(5)^2$ root lattice, we can define $\mathbb{Z}_4$ basis transformation for the metric.
The action of $Q$ on $SU(2)^2 \times SO(5)^2$ root lattice is defined as
\begin{equation}
    \begin{aligned}
     Q(e_1) &= e_1 + 2 e_2, & Q(e_2) &= -e_1 - e_2, & Q(e_3) &= e_3 + 2 e_4, \\
     Q(e_4) &= - e_3 - e_4, & Q(e_5) &= - e_5, & Q(e_6) &= - e_6.
    \end{aligned}
\end{equation}
The matrix representation of $Q$ is defined by $Q(e_i) = e_j Q_{ji}$ and the explicit matrix is shown below:
\begin{align}
    Q = \begin{pmatrix}
        1 & -1 & 0 & 0 & 0 & 0 \\
        2 & -1 & 0 & 0 & 0 & 0 \\
        0 & 0 & 1 & -1 & 0 & 0 \\
        0 & 0 & 2 & -1 & 0 & 0 \\
        0 & 0 & 0 & 0 & -1 & 0 \\
        0 & 0 & 0 & 0 & 0 & -1 \\
    \end{pmatrix},
\end{align}
where $Q^4 = 1$.

To construct the cycles on the orbifold, we apply this action $\sum \Gamma$ to 20 real three-forms.
By choosing only the linearly independent cycles from the obtained invariant cycles, the following four cycles are derived;
\begin{align}
    \begin{aligned}
    \mathbf{1}_{A^0} &\equiv \frac{1}{2} \sum \Gamma(\beta^3), \\
    \mathbf{1}_{A^1} &\equiv - \sum \Gamma(\alpha_3), \\
    \mathbf{1}_{B_0} &\equiv \frac{1}{2} \sum \Gamma(\beta^0), \\
    \mathbf{1}_{B_1} &\equiv - \sum \Gamma(\alpha_0) .
    \end{aligned}
\end{align}
Here, we also consider the sum of orbits associated with the Coxeter element and the lengths of four orbits are
\begin{align}
    |\Gamma(\alpha_0)| = 2, \quad |\Gamma(\alpha_3)| = 2, \quad |\Gamma(\beta^0)| = 2, \quad |\Gamma(\beta^3)| = 2.
\end{align}
In terms of 20 real three-forms, these cycles can be expressed as follows:
\begin{align}
    \begin{aligned}
    \mathbf{1}_{A^0} &= 2 \alpha_0 - \alpha_1 - \alpha_2, \\
    \mathbf{1}_{A^1} &= \beta^0 + \beta^1 + \beta^2, \\
    \mathbf{1}_{B_0} &= - 2 \alpha_3 - \beta^1 - \beta^2, \\
    \mathbf{1}_{B_1} &= - \alpha_1 - \alpha_2 - \beta^3.
    \end{aligned}
\end{align}
On $T^6/\mathbb{Z}_4$, these intersections of the dual cycles in regards to the orbifold satisfy the following relations;
\begin{align}
    \int_{T^6/\mathbb{Z}_4} \mathbf{1}_{A^0} \wedge \mathbf{1}_{B_0} =2, \quad \int_{T^6/\mathbb{Z}_4} \mathbf{1}_{A^1} \wedge \mathbf{1}_{B_1} = 2,
\end{align}
where the other intersection numbers are 0.

Considering the discussion in section \ref{sec:holomorphic_three-form}, we can construct the complex one-form as follows:
\begin{align}
    \begin{alignedat}{2}
    dz^1 &= dx^1 - \left( \frac{1}{2} - \frac{i}{2} \right) dx^2, \\
    dz^2 &= dx^3 - \left( \frac{1}{2} - \frac{i}{2} \right) dx^4, \\
    dz^3 &= dx^5  + U dx^6.
    \end{alignedat}
\end{align}
Consequently, by using these complex one-forms, the holomorphic three-form is defined as $\Omega = dz^1 \wedge dz^2 \wedge dz^3$ where $\int_{T^6/\mathbb{Z}_4} \Omega \wedge \bar{\Omega} = 2 i \text{Im} U$.
Moreover, the period vector is denoted by
\begin{align}
    \begin{aligned}
        \Pi &\equiv \begin{pmatrix}
               X^0 \\
               X^1 \\
               F_0 \\
               F_1
               \end{pmatrix}
            &= \begin{pmatrix}
               \frac{1}{\sqrt{2}} \int_{T^6/\mathbb{Z}_4} \Omega \wedge \mathbf{1}_{B_0} \\
               \frac{1}{\sqrt{2}} \int_{T^6/\mathbb{Z}_4} \Omega \wedge \mathbf{1}_{B_1} \\
               \frac{1}{\sqrt{2}} \int_{T^6/\mathbb{Z}_4} \Omega \wedge \mathbf{1}_{A^0} \\
               \frac{1}{\sqrt{2}} \int_{T^6/\mathbb{Z}_4} \Omega \wedge \mathbf{1}_{A^1}
               \end{pmatrix} 
            &= \begin{pmatrix}
                \frac{1}{\sqrt{2}}\\
                - \frac{i U}{\sqrt{2}} \\
                \frac{U}{\sqrt{2}} \\
                \frac{i}{\sqrt{2}}
               \end{pmatrix},
    \end{aligned}
\end{align}
which satisfies $\Pi^{\dagger} \Sigma \Pi = 2 i \text{Im} U$.
Since the overall factor can be absorbed by K\"{a}ler transformation, we consider a modified period vector $\Pi' = \left(1, -i U, U, i \right)^t$ in the discussion of modular transformation.

Here, we consider the quantization which has no contribution of the twisted sector regarding $h^{2,1}$.
By using these definitions, the three-form fluxes on the cycles which are invariant under the Coxeter element $Q$ can be denoted by
\begin{align}
    \begin{alignedat}{2}
        \frac{1}{2} \int_{T^6/\mathbb{Z}_4} (F_3, H_3) \wedge \mathbf{1}_{A^0} &= \frac{1}{2} (a^3, c^3), \\
        \frac{1}{2} \int_{T^6/\mathbb{Z}_4} (F_3, H_3) \wedge \mathbf{1}_{A^1} &= - (- b_3, - d_3), \\
        \frac{1}{2} \int_{T^6/\mathbb{Z}_4} (F_3, H_3) \wedge \mathbf{1}_{B_0} &= \frac{1}{2} (a^0, c^0), \\
        \frac{1}{2} \int_{T^6/\mathbb{Z}_4} (F_3, H_3) \wedge \mathbf{1}_{B_1} &= - (- b_0, - d_0).
    \end{alignedat}
\end{align}
Taking account of the quantization of these three-form fluxes, it is necessary to choose specific constant multiples for the flux quanta as follows:
\begin{align}
        (a^0, c^0) \in 4 \mathbb{Z},\quad (a^3, c^3) \in 4 \mathbb{Z},\quad (b_0, d_0) \in 2 \mathbb{Z},\quad  (b_3, d_3) \in 2 \mathbb{Z}.
\end{align}
By using these results, we can construct the expansion of $G_3$ as 
\begin{align}
   \begin{aligned}
        G_3 =~& \frac{1}{2} (a^0 - S c^0) \mathbf{1}_{A^0} + (b_0 - S d_0) \mathbf{1}_{A^1} - \frac{1}{2} (a^3 - S c^3) \mathbf{1}_{B_0} - (b_3 - S d_3) \mathbf{1}_{B_1}.
   \end{aligned}
\end{align}
Then, we can describe $N_{\rm flux}$ as follows:
\begin{align}
    \begin{aligned}
         N_{\rm flux} &= \int H_3 \wedge F_3 = \frac{1}{2} (- a^3 c^0 + a^0 c^3 - 4 b_3 d_0 + 4 b_0 d_3) \in 8 \mathbb{Z}.
    \end{aligned}
\end{align}
Since we can obtain the three-form fluxes and the holomorphic three-form, the superpotential can be denoted by them.
$A, B, C$ and $D$ in the case of $T^6/\mathbb{Z}_4$ orientifold with $SU(2)^2 \times SO(5)^2$ root lattice can be found as follows:
\begin{equation}
    \begin{alignedat}{5}
        A &= \frac{a^3}{2} - i b_0, \quad & C &= - \frac{a^0}{2} - i b_3, \\
        B &= - \frac{c_3}{2} + i d_0, \quad & D &= \frac{c_0}{2} + i d_3.
    \end{alignedat}
\end{equation}

\subsection{$T^6/\mathbb{Z}_{6-\rm \greekii}$ orbifold with $SU(3) \times SO(8)$ root lattice}

Considering $T^6/\mathbb{Z}_{6-\rm \greekii}$ orbifold, it is required that the six-dimensional metric is invariant under the action of $\Gamma = \mathbb{Z}_6$.
By using the Coxeter element $Q$ corresponding to $SU(3) \times SO(8)$ root lattice, we can define $\mathbb{Z}_6$ basis transformation for the metric.
The action of $Q$ on $SU(3) \times SO(8)$ root lattice is defined as
\begin{equation}
    \begin{aligned}
     Q(e_1) &= e_2, & Q(e_2) &= e_1 + e_2 + e_3 + e_4, \\
     Q(e_3) &= - e_1 - e_2 - e_3, & Q(e_4) &= - e_1 - e_2 - e_4, \\
     Q(e_5) &= e_6, & Q(e_6) &= - e_5 - e_6.
    \end{aligned}
\end{equation}
The matrix representation of $Q$ is defined by $Q(e_i) = e_j Q_{ji}$ and the explicit matrix is shown below:
\begin{align}
    Q = \begin{pmatrix}
        0 & 1 & -1 & -1 & 0 & 0 \\
        1 & 1 & -1 & -1 & 0 & 0 \\
        0 & 1 & -1 & 0 & 0 & 0 \\
        0 & 1 & -1 & -1 & 0 & 0 \\
        0 & 0 & 0 & 0 & 0 & -1 \\
        0 & 0 & 0 & 0 & 1 & -1 \\
    \end{pmatrix},
\end{align}
where $Q^6 = 1$.

To construct the cycles on the orbifold, we apply this action $\sum \Gamma$ to 20 real three-forms.
By choosing only the linearly independent cycles from the obtained invariant cycles, the following four cycles are derived;
\begin{align}
    \begin{aligned}
    \mathbf{1}_{A^0} &\equiv - \sum \Gamma(\beta^0) - 2 \sum \Gamma(\beta^3), \\
    \mathbf{1}_{A^1} &\equiv 2 \sum \Gamma(\beta^0) - \sum \Gamma(\beta^2), \\
    \mathbf{1}_{B_0} &\equiv - 2 \sum \Gamma(\beta^1) - \sum \Gamma(\beta^2),\\
    \mathbf{1}_{B_1} &\equiv - \sum \Gamma(\beta^0) .
    \end{aligned}
\end{align}
Here, we also consider the sum of orbits associated with the Coxeter element and the lengths of four orbits are
\begin{align}
    |\Gamma(\beta^0)| = 3, \quad |\Gamma(\beta^1)| = 3, \quad |\Gamma(\beta^2)| = 3, \quad |\Gamma(\beta^3)| = 3.
\end{align}
In terms of 20 real three-forms, these cycles can be expressed as follows:
\begin{align}
    \begin{aligned}
    \mathbf{1}_{A^0} &= - \alpha_1 - \beta^0 - 3 \beta^1 + \beta^2 - 2 \beta^3 - \gamma_2 + 3 \delta^2 + \delta^4, \\
    \mathbf{1}_{A^1} &= - 2 \alpha_0 +2 \alpha_1 -4 \alpha_2 + \alpha_3  + \beta^0 - 2 \beta^1 + \beta^2 + \beta^3 + 3 \gamma_2 + 2 \gamma_4 + \delta^2 + 3 \delta^4, \\
    \mathbf{1}_{B_0} &= \alpha_3 - \beta^0 -2 \beta^1 + \beta^2 - \beta^3 - \gamma_2 + \delta^2 + \delta^4, \\
    \mathbf{1}_{B_1} &= - \alpha_1 + 2 \alpha_2 - \beta^0 + \beta^1 - \beta^2 - 2 \gamma_4 - \delta^2 - \delta^4.
    \end{aligned}
\end{align}
On $T^6/\mathbb{Z}_{6-\rm \greekii}$, these intersections of the dual cycles in regards to the orbifold satisfy the following relations;
\begin{align}
    \int_{T^6/\mathbb{Z}_{6-\rm \greekii}} \mathbf{1}_{A^0} \wedge \mathbf{1}_{B_0} = 6, \quad \int_{T^6/\mathbb{Z}_{6-\rm \greekii}} \mathbf{1}_{A^1} \wedge \mathbf{1}_{B_1} = 6,
\end{align}
where the other intersection numbers are 0.

Considering the discussion in section \ref{sec:holomorphic_three-form}, we can construct the complex one-form as follows:
\begin{align}
    \begin{alignedat}{2}
    dz^1 &= dx^1 + e^{2 \pi i / 6} dx^2 - dx^3 - dx^4, \\
    dz^2 &= dx^5 + e^{2 \pi i / 3} dx^6, \\
    dz^3 &= \frac{1}{3} \left[ dx^1 - dx^2 + dx^3 + U (dx^1 - dx^2 + dx^4) \right].
    \end{alignedat}
\end{align}
Consequently, by using these complex one-forms, the holomorphic three-form is defined as $\Omega = dz^1 \wedge dz^2 \wedge dz^3$ where $\int_{T^6/\mathbb{Z}_4} \Omega \wedge \bar{\Omega} = 2 i \text{Im} U$.
Moreover, the period vector is denoted by
\begin{align}
    \begin{aligned}
        \Pi &\equiv \begin{pmatrix}
               X^0 \\
               X^1 \\
               F_0 \\
               F_1
               \end{pmatrix}
            &= \begin{pmatrix}
               \frac{1}{\sqrt{6}} \int_{T^6/\mathbb{Z}_{6-\rm \greekii}} \Omega \wedge \mathbf{1}_{B_0} \\
               \frac{1}{\sqrt{6}} \int_{T^6/\mathbb{Z}_{6-\rm \greekii}} \Omega \wedge \mathbf{1}_{B_1} \\
               \frac{1}{\sqrt{6}} \int_{T^6/\mathbb{Z}_{6-\rm \greekii}} \Omega \wedge \mathbf{1}_{A^0} \\
               \frac{1}{\sqrt{6}} \int_{T^6/\mathbb{Z}_{6-\rm \greekii}} \Omega \wedge \mathbf{1}_{A^1}
               \end{pmatrix} 
            &= \begin{pmatrix}
                - \frac{i}{\sqrt{2}} \\
                \frac{U + 2}{\sqrt{6}} \\
                \frac{i (U + 2)}{\sqrt{2}} \\
                - \sqrt{\frac{3}{2}}
               \end{pmatrix},
    \end{aligned}
\end{align}
which satisfies $\Pi^{\dagger} \Sigma \Pi = 2 i \text{Im} U$.
Since the overall factor can be absorbed by K\"{a}ler transformation, we consider a modified period vector $\Pi' = \left(1, \frac{-i U'}{\sqrt{3}}, U', i \sqrt{3} \right)^t$ in the discussion of modular transformation.
Here, we define $U'$ as $U' \equiv U + 2$.

Here, we consider the quantization which has no contribution of the twisted sector regarding $h^{2,1}$.
By using these definitions, the three-form fluxes on the cycles which are invariant under the Coxeter element $Q$ can be denoted by
\begin{align}
    \begin{alignedat}{2}
        \frac{1}{6} \int_{T^6/\mathbb{Z}_{6-\rm \greekii}} (F_3, H_3) \wedge \mathbf{1}_{A^0} &= \frac{1}{2} \left(- (a^0, c^0) - 2 (a^3, c^3) \right), \\
        \frac{1}{6} \int_{T^6/\mathbb{Z}_{6-\rm \greekii}} (F_3, H_3) \wedge \mathbf{1}_{A^1} &= \frac{1}{2} \left(2 (a^0, c^0) - (a^2, c^2) \right), \\
        \frac{1}{6} \int_{T^6/\mathbb{Z}_{6-\rm \greekii}} (F_3, H_3) \wedge \mathbf{1}_{B_0} &= \frac{1}{2} \left(- 2 (a^1, c^1) - (a^2, c^2) \right) \\
        \frac{1}{6} \int_{T^6/\mathbb{Z}_{6-\rm \greekii}} (F_3, H_3) \wedge \mathbf{1}_{B_1} &= - \frac{1}{2} (a^0, c^0).
    \end{alignedat}
\end{align}
Taking account of the quantization of these three-form fluxes, it is necessary to choose specific constant multiples for the flux quanta as follows:
\begin{align}
        (a^0, c^0) \in 4 \mathbb{Z},\quad (a^1, c^1) \in 2 \mathbb{Z},\quad (a^2, c^2) \in 4 \mathbb{Z},\quad  (a^3, c^3) \in 2 \mathbb{Z}.
\end{align}
By using these results, we can construct the expansion of $G_3$ as 
\begin{align}
   \begin{aligned}
        G_3 =~& - \left( (a^1 - S c^1) + \frac{1}{2} (a^2 - S c^2) \right) \mathbf{1}_{A^0} - \frac{1}{2} (a^0 - S c^0) \mathbf{1}_{A^1} \\
        &+ \left( \frac{1}{2} (a^0 - S c^0) + (a^3 - S c^3) \right) \mathbf{1}_{B_0} - \left( (a^0 - S c^0) + \frac{1}{2} (a^2 - S c^2) \right) \mathbf{1}_{B_1}.
   \end{aligned}
\end{align}
Then, we can describe $N_{\rm flux}$ as follows:
\begin{align}
    \begin{aligned}
         N_{\rm flux} &= \int H_3 \wedge F_3 = - 3 (a^0 c^1 + 2 a^3 c^1 + a^0 c^2 + a^3 c^2 - a^2 (c^0 + c^3) - a^1 (c^0 + 2 c^3)) \in 24 \mathbb{Z}.
    \end{aligned}
\end{align}
Since we can obtain the three-form fluxes and the holomorphic three-form, the superpotential can be denoted by them.
$A, B, C$ and $D$ in the case of $T^6/\mathbb{Z}_{6-\rm \greekii}$ orientifold with $SU(3) \times SO(8)$ root lattice can be found as follows:
\begin{equation}
    \begin{alignedat}{5}
        A &= \left( \frac{1}{2} - \frac{i \sqrt{3}}{2} \right) a^0 + 2 i \sqrt{3} a^1 + (1 + i \sqrt{3}) a^2 - i \sqrt{3} a^3, \\
         B &= \left( - \frac{1}{2} + \frac{i \sqrt{3}}{2} \right) c^0 - 2 i \sqrt{3} c^1 + (- 1 - i \sqrt{3}) c^2 + i \sqrt{3} c^3, \\
        C &= a^0 + i \sqrt{3} a^1 + \left( \frac{1}{2} + \frac{i \sqrt{3}}{2} \right) a^2, \\
        D &= - c^0 - i \sqrt{3} c^1 - \left( \frac{1}{2} + \frac{i \sqrt{3}}{2} \right) c^2.
    \end{alignedat}
\end{equation}

\subsection{$T^6/\mathbb{Z}_{6-\rm \greekii}$ orbifold with $SU(2)^2 \times SU(3)^2$ root lattice}

Considering $T^6/\mathbb{Z}_{6-\rm \greekii}$ orbifold, it is required that the six-dimensional metric is invariant under the action of $\Gamma = \mathbb{Z}_6$.
By using the Coxeter element $Q$ corresponding to $SU(2)^2 \times SU(3)^2$ root lattice, we can define $\mathbb{Z}_4$ basis transformation for the metric.
The action of $Q$ on $SU(2)^2 \times SU(3)^2$ root lattice is defined as
\begin{equation}
    \begin{aligned}
     Q(e_1) &= - e_1, & Q(e_2) &= - e_2, \\ 
     Q(e_3) &= - e_5, & Q(e_4) &= - e_5 + e_6, \\
     Q(e_5) &= e_4, & Q(e_6) &= e_3.
    \end{aligned}
\end{equation}
The matrix representation of $Q$ is defined by $Q(e_i) = e_j Q_{ji}$ and the explicit matrix is shown below:
\begin{align}
    Q = \begin{pmatrix}
        -1 & 0 & 0 & 0 & 0 & 0 \\
        0 & -1 & 0 & 0 & 0 & 0 \\
        0 & 0 & 0 & 0 & 0 & 1 \\
        0 & 0 & 0 & 0 & 1 & 0 \\
        0 & 0 & -1 & -1 & 0 & 0 \\
        0 & 0 & 0 & 1 & 0 & 0 \\
    \end{pmatrix},
\end{align}
where $Q^6 = 1$.

To construct the cycles on the orbifold, we apply this action $\sum \Gamma$ to 20 real three-forms.
By choosing only the linearly independent cycles from the obtained invariant cycles, the following four cycles are derived;
\begin{align}
    \begin{aligned}
    \mathbf{1}_{A^0} &\equiv \sum \Gamma(\alpha_0), \\
    \mathbf{1}_{A^1} &\equiv \sum \Gamma(\delta^3), \\
    \mathbf{1}_{B_0} &\equiv \sum \Gamma(\beta^0), \\
    \mathbf{1}_{B_1} &\equiv - \sum \Gamma(\gamma_3) .
    \end{aligned}
\end{align}
Here, we also consider the sum of orbits associated with the Coxeter element and the lengths of four orbits are
\begin{align}
    |\Gamma(\alpha_0)| = 3, \quad |\Gamma(\beta^0)| = 6, \quad |\Gamma(\gamma_3)| = 2, \quad |\Gamma(\delta^3)| = 2.
\end{align}
In terms of 20 real three-forms, these cycles can be expressed as follows:
\begin{align}
    \begin{aligned}
    \mathbf{1}_{A^0} &= \alpha_0 - \alpha_2 - \alpha_3 + 2 \beta^1, \\
    \mathbf{1}_{A^1} &= \delta^3 + \delta^5, \\
    \mathbf{1}_{B_0} &= - \alpha_1 + 2 \beta^0 - \beta^2 - \beta^3, \\
    \mathbf{1}_{B_1} &= - \gamma_3 - \gamma_5.
    \end{aligned}
\end{align}
On $T^6/\mathbb{Z}_{6-\rm \greekii}$, these intersections of the dual cycles in regards to the orbifold satisfy the following relations;
\begin{align}
    \int_{T^6/\mathbb{Z}_{6-\rm \greekii}} \mathbf{1}_{A^0} \wedge \mathbf{1}_{B_0} = 6, \quad \int_{T^6/\mathbb{Z}_{6-\rm \greekii}} \mathbf{1}_{A^1} \wedge \mathbf{1}_{B_1} = 2,
\end{align}
where the other intersection numbers are 0.

Considering the discussion in section \ref{sec:holomorphic_three-form}, we can construct the complex one-form as follows:
\begin{align}
    \begin{alignedat}{2}
    dz^1 &= \frac{1}{\sqrt{3}} \left( dx^3 + e^{- 2 \pi i / 6} dx^4 + e^{- 2 \pi i / 3} dx^5 + e^{- 2 \pi i / 6} dx^6 \right), \\
    dz^2 &= \frac{1}{2} \left( dx^3 - e^{- 2 \pi i / 3} dx^4 + e^{- 2 \pi i / 6} dx^5 + e^{- 2 \pi i / 3} dx^6 \right), \\
    dz^3 &= dx^1 + U dx^2.
    \end{alignedat}
\end{align}
Consequently, by using these complex one-forms, the holomorphic three-form is defined as $\Omega = dz^1 \wedge dz^2 \wedge dz^3$ where $\int_{T^6/\mathbb{Z}_{6-\rm \greekii}} \Omega \wedge \bar{\Omega} = 2 i \text{Im} U$.
Moreover, the period vector is denoted by
\begin{align}
    \begin{aligned}
        \Pi &\equiv \begin{pmatrix}
               X^0 \\
               X^1 \\
               F_0 \\
               F_1
               \end{pmatrix}
            &= \begin{pmatrix}
               \frac{1}{\sqrt{6}} \int_{T^6/\mathbb{Z}_{6-\rm \greekii}} \Omega \wedge \mathbf{1}_{B_0} \\
               \frac{1}{\sqrt{2}} \int_{T^6/\mathbb{Z}_{6-\rm \greekii}} \Omega \wedge \mathbf{1}_{B_1} \\
               \frac{1}{\sqrt{6}} \int_{T^6/\mathbb{Z}_{6-\rm \greekii}} \Omega \wedge \mathbf{1}_{A^0} \\
               \frac{1}{\sqrt{2}} \int_{T^6/\mathbb{Z}_{6-\rm \greekii}} \Omega \wedge \mathbf{1}_{A^1}
               \end{pmatrix} 
            &= \begin{pmatrix}
                \frac{1}{\sqrt{2}}\\
                - \frac{i U}{\sqrt{2}} \\
                \frac{U}{\sqrt{2}} \\
                \frac{i}{\sqrt{2}}
               \end{pmatrix},
    \end{aligned}
\end{align}
which satisfies $\Pi^{\dagger} \Sigma \Pi = 2 i \text{Im} U$.
Since the overall factor can be absorbed by K\"{a}ler transformation, we consider a modified period vector $\Pi' = \left(1, -i U, U', i \right)^t$ in the discussion of modular transformation.

Here, we consider the quantization which has no contribution of the twisted sector regarding $h^{2,1}$.
By using these definitions, the three-form fluxes on the cycles which are invariant under the Coxeter element $Q$ can be denoted by
\begin{align}
    \begin{alignedat}{2}
        \frac{1}{6} \int_{T^6/\mathbb{Z}_{6-\rm \greekii}} (F_3, H_3) \wedge \mathbf{1}_{A^0} &= \frac{1}{2} (- b_0, - d_0), \\
        \frac{1}{2} \int_{T^6/\mathbb{Z}_{6-\rm \greekii}} (F_3, H_3) \wedge \mathbf{1}_{A^1} &= (e^3, g^3), \\
        \frac{1}{6} \int_{T^6/\mathbb{Z}_{6-\rm \greekii}} (F_3, H_3) \wedge \mathbf{1}_{B_0} &= (a^0, c^0), \\
        \frac{1}{2} \int_{T^6/\mathbb{Z}_{6-\rm \greekii}} (F_3, H_3) \wedge \mathbf{1}_{B_1} &= (- f_3, - h_3).
    \end{alignedat}
\end{align}
Taking account of the quantization of these three-form fluxes, it is necessary to choose specific constant multiples for the flux quanta as follows:
\begin{align}
        (a^0, c^0) \in 2 \mathbb{Z},\quad (b_0, d_0) \in 4 \mathbb{Z},\quad (e^3, g^3) \in 2 \mathbb{Z},\quad  (f_3, h_3) \in 2 \mathbb{Z}.
\end{align}
By using these results, we can construct the expansion of $G_3$ as 
\begin{align}
   \begin{aligned}
        G_3 =~& (a^0 - S c^0) \mathbf{1}_{A^0} - (f_3 - S h_3) \mathbf{1}_{A^1} + \frac{1}{2} (b_0 - S d_0) \mathbf{1}_{B_0} - (e^3 - S g^3) \mathbf{1}_{B_1}.
   \end{aligned}
\end{align}
Then, we can describe $N_{\rm flux}$ as follows:
\begin{align}
    \begin{aligned}
         N_{\rm flux} &= \int H_3 \wedge F_3 = 3 b_0 c^0 - 3 a^0 d_0 -2 f_3 g^3 + 2 e^3 h_3 \in 8 \mathbb{Z}.
    \end{aligned}
\end{align}
Since we can obtain the three-form fluxes and the holomorphic three-form, the superpotential can be denoted by them.
$A, B, C$ and $D$ in the case of $T^6/\mathbb{Z}_{6-\rm \greekii}$ orientifold with $SU(2)^2 \times SU(3)^2$ root lattice can be found as follows:
\begin{equation}
    \begin{alignedat}{5}
        A &= - \frac{\sqrt{3}}{2} b_0 + i f_3, \quad & C &= - \sqrt{3} a^0 - i e^3, \\
        B &= \frac{\sqrt{3}}{2} d_0 - i h_3, \quad & D &= \sqrt{3} c^0 + i g^3.
    \end{alignedat}
\end{equation}

\subsection{$T^6/\mathbb{Z}_{6-\rm \greekii}$ orbifold with $SU(2)^2 \times SU(3) \times G_2$ root lattice}

Considering $T^6/\mathbb{Z}_{6-\rm \greekii}$ orbifold, it is required that the six-dimensional metric is invariant under the action of $\Gamma = \mathbb{Z}_6$.
By using the Coxeter element $Q$ corresponding to $SU(2)^2 \times SU(3) \times G_2$ root lattice, we can define $\mathbb{Z}_6$ basis transformation for the metric.
The action of $Q$ on $SU(2)^2 \times SU(3) \times G_2$ root lattice is defined as
\begin{equation}
    \begin{aligned}
     Q(e_1) &= 2 e_1 + 3 e_2, & Q(e_2) &= - e_1 - e_2, \\ 
     Q(e_3) &= e_4, & Q(e_4) &= - e_3 - e_6, \\
     Q(e_5) &= - e_5, & Q(e_6) &= - e_6.
    \end{aligned}
\end{equation}
The matrix representation of $Q$ is defined by $Q(e_i) = e_j Q_{ji}$ and the explicit matrix is shown below:
\begin{align}
    Q = \begin{pmatrix}
        2 & -1 & 0 & 0 & 0 & 0 \\
        3 & -1 & 0 & 0 & 0 & 0 \\
        0 & 0 & 0 & -1 & 0 & 0 \\
        0 & 0 & 1 & -1 & 0 & 0 \\
        0 & 0 & 0 & 0 & -1 & 0 \\
        0 & 0 & 0 & 0 & 0 & -1 \\
    \end{pmatrix},
\end{align}
where $Q^6 = 1$.

To construct the cycles on the orbifold, we apply this action $\sum \Gamma$ to 20 real three-forms.
By choosing only the linearly independent cycles from the obtained invariant cycles, the following four cycles are derived;
\begin{align}
    \begin{aligned}
    \mathbf{1}_{A^0} &\equiv \frac{1}{3} \sum \Gamma(\beta^3), \\
    \mathbf{1}_{A^1} &\equiv - \sum \Gamma(\alpha_3), \\
    \mathbf{1}_{B_0} &\equiv \frac{1}{3} \sum \Gamma(\beta^0), \\
    \mathbf{1}_{B_1} &\equiv - \sum \Gamma(\alpha_0) .
    \end{aligned}
\end{align}
Here, we also consider the sum of orbits associated with the Coxeter element and the lengths of four orbits are
\begin{align}
    |\Gamma(\alpha_0)| = 3, \quad |\Gamma(\alpha_3)| = 3, \quad |\Gamma(\beta^0)| = 3, \quad |\Gamma(\beta^3)| = 3.
\end{align}
In terms of 20 real three-forms, these cycles can be expressed as follows:
\begin{align}
    \begin{aligned}
    \mathbf{1}_{A^0} &= 2 \alpha_0 - \alpha_1 - \alpha_2, \\
    \mathbf{1}_{A^1} &= 2 \beta^0 + 3 \beta^1 + \beta^2, \\
    \mathbf{1}_{B_0} &= - 2 \alpha_3 - \beta^1 - \beta^2, \\
    \mathbf{1}_{B_1} &= - \alpha_1 - 3 \alpha_2 - 2 \beta^3.
    \end{aligned}
\end{align}
On $T^6/\mathbb{Z}_{6-\rm \greekii}$, these intersections of the dual cycles in regards to the orbifold satisfy the following relations;
\begin{align}
    \int_{T^6/\mathbb{Z}_{6-\rm \greekii}} \mathbf{1}_{A^0} \wedge \mathbf{1}_{B_0} = 2, \quad \int_{T^6/\mathbb{Z}_{6-\rm \greekii}} \mathbf{1}_{A^1} \wedge \mathbf{1}_{B_1} = 6,
\end{align}
where the other intersection numbers are 0.

Considering the discussion in section \ref{sec:holomorphic_three-form}, we can construct the complex one-form as follows:
\begin{align}
    \begin{alignedat}{2}
    dz^1 &= dx^1 + \frac{1}{\sqrt{3}}e^{5 \pi i / 6} dx^2, \\
    dz^2 &= dx^3 + e^{2 \pi i / 3} dx^4, \\
    dz^3 &= dx^5 + U dx^6.
    \end{alignedat}
\end{align}
Consequently, by using these complex one-forms, the holomorphic three-form is defined as $\Omega = dz^1 \wedge dz^2 \wedge dz^3$ where $\int_{T^6/\mathbb{Z}_{6-\rm \greekii}} \Omega \wedge \bar{\Omega} = 2 i \text{Im} U$.
Moreover, the period vector is denoted by
\begin{align}
    \begin{aligned}
        \Pi &\equiv \begin{pmatrix}
               X^0 \\
               X^1 \\
               F_0 \\
               F_1
               \end{pmatrix}
            &= \begin{pmatrix}
               \frac{1}{\sqrt{2}} \int_{T^6/\mathbb{Z}_{6-\rm \greekii}} \Omega \wedge \mathbf{1}_{B_0} \\
               \frac{1}{\sqrt{6}} \int_{T^6/\mathbb{Z}_{6-\rm \greekii}} \Omega \wedge \mathbf{1}_{B_1} \\
               \frac{1}{\sqrt{2}} \int_{T^6/\mathbb{Z}_{6-\rm \greekii}} \Omega \wedge \mathbf{1}_{A^0} \\
               \frac{1}{\sqrt{6}} \int_{T^6/\mathbb{Z}_{6-\rm \greekii}} \Omega \wedge \mathbf{1}_{A^1}
               \end{pmatrix} 
            &= \begin{pmatrix}
                \frac{1}{\sqrt{2}}\\
                - \frac{i U}{\sqrt{2}} \\
                \frac{U}{\sqrt{2}} \\
                \frac{i}{\sqrt{2}}
               \end{pmatrix},
    \end{aligned}
\end{align}
which satisfies $\Pi^{\dagger} \Sigma \Pi = 2 i \text{Im} U$.
Since the overall factor can be absorbed by K\"{a}ler transformation, we consider a modified period vector $\Pi' = \left(1, -i U, U, i \right)^t$ in the discussion of modular transformation.

Here, we consider the quantization which has no contribution of the twisted sector regarding $h^{2,1}$.
By using these definitions, the three-form fluxes on the cycles which are invariant under the Coxeter element $Q$ can be denoted by
\begin{align}
    \begin{alignedat}{2}
        \frac{1}{2} \int_{T^6/\mathbb{Z}_{6-\rm \greekii}} (F_3, H_3) \wedge \mathbf{1}_{A^0} &= \frac{1}{2} (a^3, c^3), \\
        \frac{1}{6} \int_{T^6/\mathbb{Z}_{6-\rm \greekii}} (F_3, H_3) \wedge \mathbf{1}_{A^1} &= - \frac{1}{2} (- b_3, - d_3), \\
        \frac{1}{2} \int_{T^6/\mathbb{Z}_{6-\rm \greekii}} (F_3, H_3) \wedge \mathbf{1}_{B_0} &= \frac{1}{2} (a^0, c^0), \\
        \frac{1}{6} \int_{T^6/\mathbb{Z}_{6-\rm \greekii}} (F_3, H_3) \wedge \mathbf{1}_{B_1} &= - \frac{1}{2} (- b_0, - d_0).
    \end{alignedat}
\end{align}
Taking account of the quantization of these three-form fluxes, it is necessary to choose specific constant multiples for the flux quanta as follows:
\begin{align}
        (a^0, c^0) \in 4 \mathbb{Z},\quad (a^3, c^3) \in 4 \mathbb{Z},\quad (b_0, d_0) \in 4 \mathbb{Z},\quad  (b_3, d_3) \in 4 \mathbb{Z}.
\end{align}
By using these results, we can construct the expansion of $G_3$ as 
\begin{align}
   \begin{aligned}
        G_3 =~& \frac{1}{2} (a^0 - S c^0) \mathbf{1}_{A^0} + \frac{1}{2} (b_0 - S d_0) \mathbf{1}_{A^1} - \frac{1}{2} (a^3 - S c^3) \mathbf{1}_{B_0} - \frac{1}{2} (b_3 - S d_3) \mathbf{1}_{B_1}.
   \end{aligned}
\end{align}
Then, we can describe $N_{\rm flux}$ as follows:
\begin{align}
    \begin{aligned}
         N_{\rm flux} &= \int H_3 \wedge F_3 = \frac{1}{2} (- a^3 c^0 + a^0 c^3 - 3 b_3 d_0 + 3 b_0 d_3) \in 8 \mathbb{Z}.
    \end{aligned}
\end{align}
Since we can obtain the three-form fluxes and the holomorphic three-form, the superpotential can be denoted by them.
$A, B, C$ and $D$ in the case of $T^6/\mathbb{Z}_{6-\rm \greekii}$ orientifold with $SU(2)^2 \times SU(3) \times G_2$ root lattice can be found as follows:
\begin{equation}
    \begin{alignedat}{5}
        A &= \frac{a^3}{2} - \frac{i \sqrt{3}}{2} b_0, \quad & C &= - \frac{a^0}{2} - \frac{i \sqrt{3}}{2} b_3, \\
        B &= - \frac{c^3}{2} + \frac{i \sqrt{3}}{2} d_0, \quad & D &= \frac{c^0}{2} + \frac{i \sqrt{3}}{2} d_3.
    \end{alignedat}
\end{equation}

\subsection{$T^6/\mathbb{Z}_{8-\rm \greekii}$ orbifold with $SU(2) \times SO(10)$ root lattice}

Considering $T^6/\mathbb{Z}_{8-\rm \greekii}$ orbifold, it is required that the six-dimensional metric is invariant under the action of $\Gamma = \mathbb{Z}_8$.
By using the Coxeter element $Q$ corresponding to $SU(2) \times SO(10)$ root lattice, we can define $\mathbb{Z}_{8-\rm \greekii}$ basis transformation for the metric.
The action of $Q$ on $SU(2) \times SO(10)$ root lattice is defined as
\begin{equation}
    \begin{aligned}
     Q(e_1) &= e_2, & Q(e_2) &= e_3, & Q(e_3) &= e_1 + e_2 + e_3 + e_4 + e_5, \\ Q(e_4) &= - e_1 - e_2 - e_3 - e_4, & Q(e_5) &= - e_1 - e_2 - e_3 - e_5, & Q(e_6) &= - e_6.
    \end{aligned}
\end{equation}
The matrix representation of $Q$ is defined by $Q(e_i) = e_j Q_{ji}$ and the explicit matrix is shown below:
\begin{align}
    Q = \begin{pmatrix}
        0 & 0 & 1 & -1 & -1 & 0 \\
        1 & 0 & 1 & -1 & -1 & 0 \\
        0 & 1 & 1 & -1 & -1 & 0 \\
        0 & 0 & 1 & -1 & 0 & 0 \\
        0 & 0 & 1 & 0 & -1 & 0 \\
        0 & 0 & 0 & 0 & 0 & -1 \\
    \end{pmatrix},
\end{align}
where $Q^8 = 1$.

To construct the cycles on the orbifold, we apply this action $\sum \Gamma$ to 20 real three-forms.
By choosing only the linearly independent cycles from the obtained invariant cycles, the following four cycles are derived;
\begin{align}
    \begin{aligned}
    \mathbf{1}_{A^0} &\equiv \sum \Gamma(\alpha_2), \\
    \mathbf{1}_{A^1} &\equiv - \frac{1}{2} \sum \Gamma(\beta^0), \\
    \mathbf{1}_{B_0} &\equiv \frac{1}{2} \sum \Gamma(\beta^1), \\
    \mathbf{1}_{B_1} &\equiv \sum \Gamma(\beta^3) .
    \end{aligned}
\end{align}
Here, we also consider the sum of orbits associated with the Coxeter element and the lengths of four orbits are
\begin{align}
    |\Gamma(\alpha_2)| = 4, \quad |\Gamma(\beta^0)| = 8, \quad |\Gamma(\beta^1)| = 8, \quad |\Gamma(\beta^3)| = 2.
\end{align}
In terms of 20 real three-forms, these cycles can be expressed as follows:
\begin{align}
    \begin{aligned}
    \mathbf{1}_{A^0} &= \alpha_1 + 2 \alpha_2 + 2 \beta^3 - \gamma_2 + \delta^5 - \delta^6, \\
    \mathbf{1}_{A^1} &= 2 \alpha_3 - \beta^0 + \beta^1 - \gamma_5 + \gamma_6 - \delta^2 + \delta^3, \\
    \mathbf{1}_{B_0} &= \beta^0 + \beta^1 + \beta^2 - \gamma_5 - \delta^3 - \delta^4, \\
    \mathbf{1}_{B_1} &= \alpha_0 - \alpha_2 + \beta^3 - \gamma_3 + \gamma_4.
    \end{aligned}
\end{align}
On $T^6/\mathbb{Z}_{8-\rm \greekii}$, these intersections of the dual cycles in regards to the orbifold satisfy the following relations;
\begin{align}
    \int_{T^6/\mathbb{Z}_4} \mathbf{1}_{A^0} \wedge \mathbf{1}_{B_0} = 4, \quad \int_{T^6/\mathbb{Z}_4} \mathbf{1}_{A^1} \wedge \mathbf{1}_{B_1} = 4,
\end{align}
where the other intersection numbers are 0.

Considering the discussion in section \ref{sec:holomorphic_three-form}, we can construct the complex one-form as follows:
\begin{align}
    \begin{alignedat}{2}
    dz^1 &= \frac{1}{2^{3/4}} \left[dx^1 + e^{2 \pi i / 8} dx^2 + i ~ dx^3 - \frac{1}{2} ( 1 + \sqrt{2} + i ) ( dx^4 + dx^5 ) \right], \\
    dz^2 &= \frac{1}{2^{3/4}} \left[dx^1 + e^{6 \pi i / 8} dx^2 - i ~ dx^3 + \frac{1}{2} ( -1 + \sqrt{2} + i ) ( dx^4 + dx^5 ) \right], \\
    dz^3 &= \frac{1}{2} ( -dx^4 + dx^5 + 2 U dx^6 ).
    \end{alignedat}
\end{align}
Consequently, by using these complex one-forms, the holomorphic three-form is defined as $\Omega = dz^1 \wedge dz^2 \wedge dz^3$ where $\int_{T^6/\mathbb{Z}_{8-\rm \greekii}} \Omega \wedge \bar{\Omega} = 2 i \text{Im} U$.
Moreover, the period vector is denoted by
\begin{align}
    \begin{aligned}
        \Pi &\equiv \begin{pmatrix}
               X^0 \\
               X^1 \\
               F_0 \\
               F_1
               \end{pmatrix}
            &= \begin{pmatrix}
               \frac{1}{2} \int_{T^6/\mathbb{Z}_{8-\rm \greekii}} \Omega \wedge \mathbf{1}_{B_0} \\
               \frac{1}{2} \int_{T^6/\mathbb{Z}_{8-\rm \greekii}} \Omega \wedge \mathbf{1}_{B_1} \\
               \frac{1}{2} \int_{T^6/\mathbb{Z}_{8-\rm \greekii}} \Omega \wedge \mathbf{1}_{A^0} \\
               \frac{1}{2} \int_{T^6/\mathbb{Z}_{8-\rm \greekii}} \Omega \wedge \mathbf{1}_{A^1}
               \end{pmatrix} 
            &= \begin{pmatrix}
                \frac{1}{2} \\
                - \frac{i U }{\sqrt{2}} \\
                U \\
                \frac{i}{\sqrt{2}}
               \end{pmatrix},
    \end{aligned}
\end{align}
which satisfies $\Pi^{\dagger} \Sigma \Pi = 2 i \text{Im} U$.
Since the overall factor can be absorbed by K\"{a}ler transformation, we consider a modified period vector $\Pi' = \left(1, - i \sqrt{2} U, 2 U, i \sqrt{2} \right)^t$ in the discussion of modular transformation.

Here, we consider the quantization which has no contribution of the twisted sector regarding $h^{2,1}$.
By using these definitions, the three-form fluxes on the cycles which are invariant under the Coxeter element $Q$ can be denoted by
\begin{align}
    \begin{alignedat}{2}
        \frac{1}{4} \int_{T^6/\mathbb{Z}_{8-\rm \greekii}} (F_3, H_3) \wedge \mathbf{1}_{A^0} &= (- b_2, - d_2), \\
        \frac{1}{4} \int_{T^6/\mathbb{Z}_{8-\rm \greekii}} (F_3, H_3) \wedge \mathbf{1}_{A^1} &= - (a^0, c^0), \\
        \frac{1}{4} \int_{T^6/\mathbb{Z}_{8-\rm \greekii}} (F_3, H_3) \wedge \mathbf{1}_{B_0} &= (a^1, c^1), \\
        \frac{1}{4} \int_{T^6/\mathbb{Z}_{8-\rm \greekii}} (F_3, H_3) \wedge \mathbf{1}_{B_1} &= \frac{1}{2} (a^3, c^3).
    \end{alignedat}
\end{align}
Taking account of the quantization of these three-form fluxes, it is necessary to choose specific constant multiples for the flux quanta as follows:
\begin{align}
        (a^0, c^0) \in 2 \mathbb{Z},\quad (a^1, c^1) \in 2 \mathbb{Z},\quad (a^3, c^3) \in 4 \mathbb{Z},\quad  (b_2, d_2) \in 2 \mathbb{Z}.
\end{align}
By using these results, we can construct the expansion of $G_3$ as 
\begin{align}
   \begin{aligned}
        G_3 =~& (a^1 - S c^1) \mathbf{1}_{A^0} + \frac{1}{2} (a^3 - S c^3) \mathbf{1}_{A^1} + (b_2 - S d_2) \mathbf{1}_{B_0} + (a^0 - S c^0) \mathbf{1}_{B_1}.
   \end{aligned}
\end{align}
Then, we can describe $N_{\rm flux}$ as follows:
\begin{align}
    \begin{aligned}
         N_{\rm flux} &= \int H_3 \wedge F_3 = 2 (- a^3 c^0 + a^0 c^3 + 2 b_2 c_1 - 2 a_1 d_2) \in 16 \mathbb{Z}.
    \end{aligned}
\end{align}
Since we can obtain the three-form fluxes and the holomorphic three-form, the superpotential can be denoted by them.
$A, B, C$ and $D$ in the case of $T^6/\mathbb{Z}_{8-\rm \greekii}$ orientifold with $SU(2) \times SO(10)$ root lattice can be found as follows:
\begin{equation}
    \begin{alignedat}{5}
        A &= - \frac{i a^3}{\sqrt{2}} - b_2, \quad & C &= i \sqrt{2} a^0 - 2 a^1, \\
        B &= \frac{i c^3}{\sqrt{2}} + d_2, \quad & D &= - i \sqrt{2} c^0 + 2 c^1.
    \end{alignedat}
\end{equation}

\subsection{$T^6/\mathbb{Z}_{8-\rm \greekii}$ orbifold with $SO(4) \times SO(9)$ root lattice}

Considering $T^6/\mathbb{Z}_{8-\rm \greekii}$ orbifold, it is required that the six-dimensional metric is invariant under the action of $\Gamma = \mathbb{Z}_8$.
By using the Coxeter element $Q$ corresponding to $SO(4) \times SO(9)$ root lattice, we can define $\mathbb{Z}_{8-\rm \greekii}$ basis transformation for the metric.
The action of $Q$ on $SO(4) \times SO(9)$ root lattice is defined as
\begin{equation}
    \begin{aligned}
     Q(e_1) &= e_2, & Q(e_2) &= e_3, & Q(e_3) &= e_1 + e_2 + e_3 + 2 e_4, \\ Q(e_4) &= - e_1 - e_2 - e_3 - e_4, & Q(e_5) &= - e_5, & Q(e_6) &= - e_6.
    \end{aligned}
\end{equation}
The matrix representation of $Q$ is defined by $Q(e_i) = e_j Q_{ji}$ and the explicit matrix is shown below:
\begin{align}
    Q = \begin{pmatrix}
        0 & 0 & 1 & -1 & 0 & 0 \\
        1 & 0 & 1 & -1 & 0 & 0 \\
        0 & 1 & 1 & -1 & 0 & 0 \\
        0 & 0 & 2 & -1 & 0 & 0 \\
        0 & 0 & 0 & 0 & -1 & 0 \\
        0 & 0 & 0 & 0 & 0 & -1 \\
    \end{pmatrix},
\end{align}
where $Q^8 = 1$.

To construct the cycles on the orbifold, we apply this action $\sum \Gamma$ to 20 real three-forms.
By choosing only the linearly independent cycles from the obtained invariant cycles, the following four cycles are derived;
\begin{align}
    \begin{aligned}
    \mathbf{1}_{A^0} &\equiv \frac{1}{2} \sum \Gamma(\alpha_2), \\
    \mathbf{1}_{A^1} &\equiv - \sum \Gamma(\beta^0), \\
    \mathbf{1}_{B_0} &\equiv \frac{1}{2} \sum \Gamma(\beta^1), \\
    \mathbf{1}_{B_1} &\equiv \sum \Gamma(\beta^3) .
    \end{aligned}
\end{align}
Here, we also consider the sum of orbits associated with the Coxeter element and the lengths of four orbits are
\begin{align}
    |\Gamma(\alpha_2)| = 4, \quad |\Gamma(\beta^2)| = 2, \quad |\Gamma(\beta^1)| = 4, \quad |\Gamma(\beta^3)| = 2.
\end{align}
In terms of 20 real three-forms, these cycles can be expressed as follows:
\begin{align}
    \begin{aligned}
    \mathbf{1}_{A^0} &= \alpha_1 + \alpha_2 + \beta^3 - \gamma_2, \\
    \mathbf{1}_{A^1} &= 2 \alpha_3 - \beta^0 + \beta^1 - \delta^2, \\
    \mathbf{1}_{B_0} &= \beta^0 + \beta^1 + \beta^2 - \delta^4, \\
    \mathbf{1}_{B_1} &= 2 \alpha_0 - \alpha_2 + \beta^3 + \gamma_4.
    \end{aligned}
\end{align}
On $T^6/\mathbb{Z}_{8-\rm \greekii}$, these intersections of the dual cycles in regards to the orbifold satisfy the following relations;
\begin{align}
    \int_{T^6/\mathbb{Z}_4} \mathbf{1}_{A^0} \wedge \mathbf{1}_{B_0} = 2, \quad \int_{T^6/\mathbb{Z}_4} \mathbf{1}_{A^1} \wedge \mathbf{1}_{B_1} = 4,
\end{align}
where the other intersection numbers are 0.

Considering the discussion in section \ref{sec:holomorphic_three-form}, we can construct the complex one-form as follows:
\begin{align}
    \begin{alignedat}{2}
    dz^1 &= \frac{1}{2^{3/4}} \left[dx^1 + e^{2 \pi i / 8} dx^2 + i ~ dx^3 - \frac{1}{2} ( 1 + \sqrt{2} + i ) dx^4 \right], \\
    dz^2 &= \frac{1}{2^{3/4}} \left[dx^1 + e^{6 \pi i / 8} dx^2 - i ~ dx^3 + \frac{1}{2} ( -1 + \sqrt{2} + i ) dx^4  \right], \\
    dz^3 &= dx^5 + U dx^6.
    \end{alignedat}
\end{align}
Consequently, by using these complex one-forms, the holomorphic three-form is defined as $\Omega = dz^1 \wedge dz^2 \wedge dz^3$ where $\int_{T^6/\mathbb{Z}_{8-\rm \greekii}} \Omega \wedge \bar{\Omega} = 2 i \text{Im} U$.
Moreover, the period vector is denoted by
\begin{align}
    \begin{aligned}
        \Pi &\equiv \begin{pmatrix}
               X^0 \\
               X^1 \\
               F_0 \\
               F_1
               \end{pmatrix}
            &= \begin{pmatrix}
               \frac{1}{\sqrt{2}} \int_{T^6/\mathbb{Z}_{8-\rm \greekii}} \Omega \wedge \mathbf{1}_{B_0} \\
               \frac{1}{2} \int_{T^6/\mathbb{Z}_{8-\rm \greekii}} \Omega \wedge \mathbf{1}_{B_1} \\
               \frac{1}{\sqrt{2}} \int_{T^6/\mathbb{Z}_{8-\rm \greekii}} \Omega \wedge \mathbf{1}_{A^0} \\
               \frac{1}{2} \int_{T^6/\mathbb{Z}_{8-\rm \greekii}} \Omega \wedge \mathbf{1}_{A^1}
               \end{pmatrix} 
            &= \begin{pmatrix}
                \frac{1}{\sqrt{2}}\\
                - \frac{i U}{\sqrt{2}} \\
                \frac{U}{\sqrt{2}} \\
                \frac{i}{\sqrt{2}}
               \end{pmatrix},
    \end{aligned}
\end{align}
which satisfies $\Pi^{\dagger} \Sigma \Pi = 2 i \text{Im} U$.
Since the overall factor can be absorbed by K\"{a}ler transformation, we consider a modified period vector $\Pi' = \left(1, -i U, U, i \right)^t$ in the discussion of modular transformation.

Here, we consider the quantization which has no contribution of the twisted sector regarding $h^{2,1}$.
By using these definitions, the three-form fluxes on the cycles which are invariant under the Coxeter element $Q$ can be denoted by
\begin{align}
    \begin{alignedat}{2}
        \frac{1}{2} \int_{T^6/\mathbb{Z}_{8-\rm \greekii}} (F_3, H_3) \wedge \mathbf{1}_{A^0} &= (- b_2 , - d_2), \\
        \frac{1}{4} \int_{T^6/\mathbb{Z}_{8-\rm \greekii}} (F_3, H_3) \wedge \mathbf{1}_{A^1} &= - \frac{1}{2} (a^0, c^0), \\
        \frac{1}{2} \int_{T^6/\mathbb{Z}_{8-\rm \greekii}} (F_3, H_3) \wedge \mathbf{1}_{B_0} &= (a^1, c^1), \\
        \frac{1}{4} \int_{T^6/\mathbb{Z}_{8-\rm \greekii}} (F_3, H_3) \wedge \mathbf{1}_{B_1} &= \frac{1}{2} (a^3, c^3).
    \end{alignedat}
\end{align}
Taking account of the quantization of these three-form fluxes, it is necessary to choose specific constant multiples for the flux quanta as follows:
\begin{align}
        (a^0, c^0) \in 4 \mathbb{Z},\quad (a^1, c^1) \in 2 \mathbb{Z},\quad (a^3, c^3) \in 4 \mathbb{Z},\quad (b_2 , d_2) \in 2 \mathbb{Z}.
\end{align}
By using these results, we can construct the expansion of $G_3$ as 
\begin{align}
   \begin{aligned}
        G_3 =~& (a^1 - S c^1) \mathbf{1}_{A^0} + \frac{1}{2} (a^3 - S c^3) \mathbf{1}_{A^1} + (b_2 - S d_2) \mathbf{1}_{B_0} + \frac{1}{2} (a^0 - S c^0) \mathbf{1}_{B_1}.
   \end{aligned}
\end{align}
Then, we can describe $N_{\rm flux}$ as follows:
\begin{align}
    \begin{aligned}
         N_{\rm flux} &= \int H_3 \wedge F_3 = - a^3 c^0 + a^0 c^3 + 2 b_2 c^1 - 2 a^2 d_2 \in 8 \mathbb{Z}.
    \end{aligned}
\end{align}
Since we can obtain the three-form fluxes and the holomorphic three-form, the superpotential can be denoted by them.
$A, B, C$ and $D$ in the case of $T^6/\mathbb{Z}_{8-\rm \greekii}$ orientifold with $SO(4) \times SO(9)$ root lattice can be found as follows:
\begin{equation}
    \begin{alignedat}{5}
        A &= - \frac{i a^3}{\sqrt{2}} - b_2, \quad & C &= \frac{i a^0}{\sqrt{2}} - a^1, \\
        B &= \frac{i c^3}{\sqrt{2}} + d_2, \quad & D &= - \frac{i c^0}{\sqrt{2}} + c^1.
    \end{alignedat}
\end{equation}

\subsection{$T^6/\mathbb{Z}_{12-\rm \greekii}$ orbifold with $SO(4) \times F_4$ root lattice}

Considering $T^6/\mathbb{Z}_{12-\rm \greekii}$ orbifold, it is required that the six-dimensional metric is invariant under the action of $\Gamma = \mathbb{Z}_{12}$.
By using the Coxeter element $Q$ corresponding to $SO(4) \times F_4$ root lattice, we can define $\mathbb{Z}_{12}$ basis transformation for the metric.
The action of $Q$ on $SO(4) \times F_4$ root lattice is defined as
\begin{equation}
    \begin{aligned}
     Q(e_1) &= e_2, & Q(e_2) &=  e_1 + e_2 + 2 e_3, & Q(e_3) &= e_4, \\ Q(e_4) &= - e_1 - e_2 - e_3 - e_4, & Q(e_5) &= - e_5, & Q(e_6) &= - e_6.
    \end{aligned}
\end{equation}
The matrix representation of $Q$ is defined by $Q(e_i) = e_j Q_{ji}$ and the explicit matrix is shown below:
\begin{align}
    Q = \begin{pmatrix}
        0 & 1 & 0 & -1 & 0 & 0 \\
        1 & 1 & 0 & -1 & 0 & 0 \\
        0 & 2 & 0 & -1 & 0 & 0 \\
        0 & 0 & 1 & -1 & 0 & 0 \\
        0 & 0 & 0 & 0 & -1 & 0 \\
        0 & 0 & 0 & 0 & 0 & -1 \\
    \end{pmatrix},
\end{align}
where $Q^{12} = 1$.

To construct the cycles on the orbifold, we apply this action $\sum \Gamma$ to 20 real three-forms.
By choosing only the linearly independent cycles from the obtained invariant cycles, the following four cycles are derived;
\begin{align}
    \begin{aligned}
    \mathbf{1}_{A^0} &\equiv \frac{1}{3} \sum \Gamma(\alpha_0), \\
    \mathbf{1}_{A^1} &\equiv \frac{1}{3} \sum \Gamma(\beta^2), \\
    \mathbf{1}_{B_0} &\equiv \frac{1}{3} \sum \Gamma(\beta^0), \\
    \mathbf{1}_{B_1} &\equiv \frac{1}{3} \sum \Gamma(\alpha_1) .
    \end{aligned}
\end{align}
Here, we also consider the sum of orbits associated with the Coxeter element and the lengths of four orbits are
\begin{align}
    |\Gamma(\alpha_0)| = 6, \quad |\Gamma(\alpha_1)| = 6, \quad |\Gamma(\beta^0)| = 6, \quad |\Gamma(\beta^2)| = 6.
\end{align}
In terms of 20 real three-forms, these cycles can be expressed as follows:
\begin{align}
    \begin{aligned}
    \mathbf{1}_{A^0} &= 2 \alpha_0 + 2\beta^3 - 2 \gamma_2 + \gamma_4, \\
    \mathbf{1}_{A^1} &= 2 \beta^1 - \delta^2 - 2 \delta^4, \\
    \mathbf{1}_{B_0} &= - 2 \alpha_3 + 2 \beta^0 + \delta^2 - 2 \delta^4, \\
    \mathbf{1}_{B_1} &= - 2 \alpha_2 + 2 \gamma_2 + \gamma_4.
    \end{aligned}
\end{align}
On $T^6/\mathbb{Z}_4$, these intersections of the dual cycles in regards to the orbifold satisfy the following relations;
\begin{align}
    \int_{T^6/\mathbb{Z}_{12-\rm \greekii}} \mathbf{1}_{A^0} \wedge \mathbf{1}_{B_0} = 4, \quad \int_{T^6/\mathbb{Z}_{12-\rm \greekii}} \mathbf{1}_{A^1} \wedge \mathbf{1}_{B_1} = 4,
\end{align}
where the other intersection numbers are 0.

Considering the discussion in section \ref{sec:holomorphic_three-form}, we can construct the complex one-form as follows:
\begin{align}
    \begin{alignedat}{2}
    dz^1 &= \frac{1}{3^{1/4}} \left[dx^1 + e^{2 \pi i / 12} dx^2 + \frac{1}{\sqrt{2}} \left( e^{11 \pi i / 12} dx^3 - e^{\pi i / 12} dx^4 \right) \right], \\
    dz^2 &= \frac{1}{3^{1/4}} \left[dx^1 + e^{10 \pi i / 12} dx^2 + \frac{1}{\sqrt{2}} \left( e^{-5 \pi i / 12} dx^3 + e^{5 \pi i / 12} dx^4 \right) \right], \\
    dz^3 &= dx^5 + U dx^6.
    \end{alignedat}
\end{align}
Consequently, by using these complex one-forms, the holomorphic three-form is defined as $\Omega = dz^1 \wedge dz^2 \wedge dz^3$ where $\int_{T^6/\mathbb{Z}_{12-\rm \greekii}} \Omega \wedge \bar{\Omega} = 2 i \text{Im} U$.
Moreover, the period vector is denoted by
\begin{align}
    \begin{aligned}
        \Pi &\equiv \begin{pmatrix}
               X^0 \\
               X^1 \\
               F_0 \\
               F_1
               \end{pmatrix}
            &= \begin{pmatrix}
               \frac{1}{2} \int_{T^6/\mathbb{Z}_{12-\rm \greekii}} \Omega \wedge \mathbf{1}_{B_0} \\
               \frac{1}{2} \int_{T^6/\mathbb{Z}_{12-\rm \greekii}} \Omega \wedge \mathbf{1}_{B_1} \\
               \frac{1}{2} \int_{T^6/\mathbb{Z}_{12-\rm \greekii}} \Omega \wedge \mathbf{1}_{A^0} \\
               \frac{1}{2} \int_{T^6/\mathbb{Z}_{12-\rm \greekii}} \Omega \wedge \mathbf{1}_{A^1}
               \end{pmatrix} 
            &= \begin{pmatrix}
                \frac{1 - i}{2}\\
                - \frac{1 + i}{2} U \\
                \frac{1 - i}{2} U \\
                \frac{1 + i}{2}
               \end{pmatrix},
    \end{aligned}
\end{align}
which satisfies $\Pi^{\dagger} \Sigma \Pi = 2 i \text{Im} U$.
Since the overall factor can be absorbed by K\"{a}ler transformation, we consider a modified period vector $\Pi' = \left(1, -i U, U, i \right)^t$ in the discussion of modular transformation.

Here, we consider the quantization which has no contribution of the twisted sector regarding $h^{2,1}$.
By using these definitions, the three-form fluxes on the cycles which are invariant under the Coxeter element $Q$ can be denoted by
\begin{align}
    \begin{alignedat}{2}
        \frac{1}{4} \int_{T^6/\mathbb{Z}_{12-\rm \greekii}} (F_3, H_3) \wedge \mathbf{1}_{A^0} &= \frac{1}{2} (- b_0, - d_0), \\
        \frac{1}{4} \int_{T^6/\mathbb{Z}_{12-\rm \greekii}} (F_3, H_3) \wedge \mathbf{1}_{A^1} &= \frac{1}{2} (a^2, c^2), \\
        \frac{1}{4} \int_{T^6/\mathbb{Z}_{12-\rm \greekii}} (F_3, H_3) \wedge \mathbf{1}_{B_0} &= \frac{1}{2} (a^0, c^0), \\
        \frac{1}{4} \int_{T^6/\mathbb{Z}_{12-\rm \greekii}} (F_3, H_3) \wedge \mathbf{1}_{B_1} &= \frac{1}{2} (- b_1, - d_1).
    \end{alignedat}
\end{align}
Taking account of the quantization of these three-form fluxes, it is necessary to choose specific constant multiples for the flux quanta as follows:
\begin{align}
        (a^0, c^0) \in 4 \mathbb{Z},\quad (a^2, c^2) \in 4 \mathbb{Z},\quad (b_0, d_0) \in 4 \mathbb{Z},\quad (b_1, d_1) \in 4 \mathbb{Z}.
\end{align}
By using these results, we can construct the expansion of $G_3$ as 
\begin{align}
   \begin{aligned}
        G_3 =~& \frac{1}{2} (a^0 - S c^0) \mathbf{1}_{A^0} - \frac{1}{2} (b_1 - S d_1) \mathbf{1}_{A^1} + \frac{1}{2} (b_0 - S d_0) \mathbf{1}_{B_0} - \frac{1}{2} (a^2 - S c^2) \mathbf{1}_{B_1}.
   \end{aligned}
\end{align}
Then, we can describe $N_{\rm flux}$ as follows:
\begin{align}
    \begin{aligned}
         N_{\rm flux} &= \int H_3 \wedge F_3 = b_0 c^0 - b_1 c^2 - a^0 d_0 + a^2 d_1 \in 16 \mathbb{Z}.
    \end{aligned}
\end{align}
Since we can obtain the three-form fluxes and the holomorphic three-form, the superpotential can be denoted by them.
$A, B, C$ and $D$ in the case of $T^6/\mathbb{Z}_{12-\rm \greekii}$ orientifold with $SO(4) \times F_4$ root lattice can be found as follows:
\begin{equation}
    \begin{alignedat}{5}
        A &= \left( - \frac{1}{2} + \frac{i}{2} \right) b_0 + \left( \frac{1}{2} + \frac{i}{2} \right) b_1, \quad & C &= \left( - \frac{1}{2} + \frac{i}{2} \right) a^0 - \left( \frac{1}{2} + \frac{i}{2} \right) a^2, \\
        B &= \left( \frac{1}{2} - \frac{i}{2} \right) d_0 - \left( \frac{1}{2} +\frac{i}{2} \right) d_1, \quad & D &= \left( \frac{1}{2} - \frac{i}{2} \right) c^0 + \left( \frac{1}{2} + \frac{i}{2} \right) c^2.
    \end{alignedat}
\end{equation}

\bibliography{references}{}
\bibliographystyle{JHEP}

\end{document}